\documentclass{ws-ijmpe}

\begin{document}

\markboth{Thomas J. Humanic}{Hanbury-Brown-Twiss Interferometry with Identical
Bosons in Relativistic Heavy Ion Collisions}

%
\catchline{}{}{}{}{}
%

\title{ Hanbury-Brown-Twiss Interferometry with Identical Bosons in Relativistic
Heavy Ion Collisions: Comparisons with Hadronic Scattering Models}

\author{Thomas J. Humanic}

\address{Department of Physics, The Ohio State University\\
Columbus, Ohio, 43210
\\humanic@mps.ohio-state.edu}

\maketitle


\begin{abstract}
Identical boson Hanbury-Brown-Twiss interferometry as applied to
relativistic heavy-ion collisions is reviewed. Emphasis is placed on
the use of hadronic scattering models to interpret the physical
significance of experimental results. Interferometric studies with
center-of-mass energies from $<1$ GeV/nucleon up to 5500 GeV/nucleon
are considered.
\end{abstract}

\section{Introduction}

\subsection{Scope of this Review}
In the present work the application of the Hanbury-Brown-Twiss
interferometry (HBT) technique to relativistic heavy-ion collisions
(RHI) is reviewed. The emphasis is placed on comparing identical
boson HBT experiments ranging in $\sqrt{s}$ from less than 1
GeV/nucleon to 200 GeV/nucleon with hadronic scattering models to
attempt to understand what has been learned from these studies over
the last twenty-five years. Predictions from such a model are also
given for future LHC Pb+Pb collisions at $5500$ GeV/nucleon. Since
the literature has been quite rich in this field during this period,
it has been necessary to be selective both in the experiments and
models which are presented. Thus, while in no way attempting to be
comprehensive, an attempt has been made to present results in this
work which are at least representative of the major developments in
the field. More comprehensive reviews of this field can be found in
the literature.\cite{lisa2005,heinz1999}

The review is organized as follows: the remainder of Section 1 gives
a brief discussion of the original work of Hanbury-Brown and Twiss
in developing HBT followed by the motivation for applying HBT to
RHI, Section 2 gives some practical information to help understand
how HBT is applied to RHI, Section 3 describes the hadronic
scattering models discussed in this paper and presents comparisons
of these models with results from two-boson HBT experiments, and
Section 4 gives a summary.

\subsection{Origins of HBT: Hanbury-Brown and Twiss experiments}
About fifty years ago, Hanbury-Brown and Twiss first suggested, and
then proved in a table-top experiment, that photon pairs exhibit a
second-order interference effect if detected simultaneously in two
detectors.\cite{hbt1} They applied this technique, which we now call
HBT when applied with any type of boson,\cite{cris1} to the
measurement of the angular diameter of stars using pairs of
photomultiplier tubes to detect optical photons\cite{hbt2} and pairs
of radiotelescope dishes to detect longer wavelength
photons.\cite{hbt3} Their measurement of the optical angular
diameter of the star Sirius located in the constellation Canis
Major, serves as a good example of the HBT method.\cite{hbt2} A
schematic layout which helps to demonstrate the principle of their
method is shown in Figure 1. Sirius is shown emitting two photons
with wavevectors $\mathbf{k}_i$ and $\mathbf{k}_j$ from points
$\mathbf{x}_i$ and $\mathbf{x}_j$, respectively, which are detected
in two photomultiplier tubes located at positions $\mathbf{a}_1$ and
$\mathbf{a}_2$ on the Earth.

\begin{figure}[th]
\centerline{\psfig{file=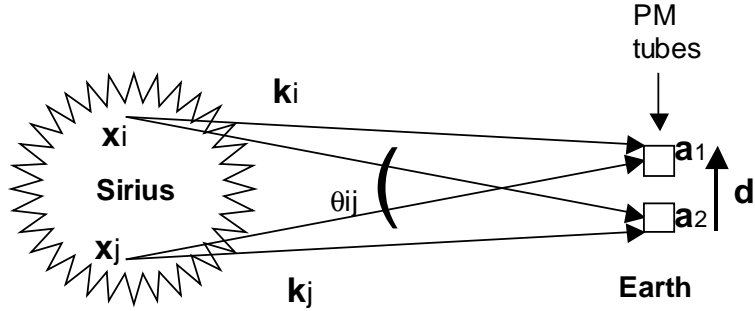,width=10cm}} \vspace*{8pt}
\caption{Schematic diagram of the method used to measure the angular
diameter of Sirius by Hanbury-Brown and Twiss in 1956.}
\end{figure}

Assuming that photons are emitted incoherently from the star at each position
and the photons are emitted as plane waves and taking a time-independent picture,
the wavefunction to detect in coincidence the two photons in the two
detectors on Earth,
$\Psi(\mathbf{k}_i,\mathbf{k}_j;\mathbf{x}_i,\mathbf{x}_j)$, is
\begin{eqnarray}
\label{hbt1}
\Psi(\mathbf{k}_i,\mathbf{k}_j;\mathbf{x}_i,\mathbf{x}_j)=
b[\exp(i\mathbf{k}_i\cdot(\mathbf{a}_1-\mathbf{x}_i))
\exp(i\mathbf{k}_j\cdot(\mathbf{a}_2-\mathbf{x}_j)) \nonumber\\ +
\exp(i\mathbf{k}_i\cdot(\mathbf{a}_2-\mathbf{x}_i))
\exp(i\mathbf{k}_j\cdot(\mathbf{a}_1-\mathbf{x}_j))]
\end{eqnarray}
where the second term arises due to the path ambiguity of detecting bosons (as
shown in Figure 1), and where $b$ is a normalization constant. The probability
to detect the two photons, $P_{ij}$, is just the square of
$\Psi$, i.e. $\Psi^*\Psi$, and thus given by
\begin{equation}
\label{hbt2} P_{ij}(\Delta\mathbf{k},\mathbf{d})=
b^2[1+\cos(\Delta\mathbf{k}\cdot\mathbf{d})]
\end{equation}
where $\Delta\mathbf{k}=\mathbf{k}_i-\mathbf{k}_j$ and
$\mathbf{d}=\mathbf{a}_1- \mathbf{a}_2$. The separation between the
two detectors, $d$, is called the baseline. It is seen in Eq.
(\ref{hbt2}) that a) the $cos$ term in $P_{ij}$ is a result of the
path ambiguity term in Eq. (\ref{hbt1}), and b) $P_{ij}$ does not
depend on the photon emission positions at the star but only on the
differences between the wavevectors and the detector positions. One
can write Eq. (\ref{hbt2}) in a more convenient form with the
approximations $k_i\simeq
 k_j\equiv k$,
$\Delta\mathbf{k}\cdot\mathbf{d}\simeq\mid\Delta\mathbf{k}\mid$,
$\mid\Delta\mathbf{k}\mid\simeq\theta_{ij}k=2\pi\theta_{ij}/\lambda$,
taking $b=1$, and defining the correlator, $C_{ij}(d)\equiv
P_{ij}-1$, resulting in
\begin{equation}
\label{hbt3}
C_{ij}(d)=\cos(2\pi\theta_{ij}d/\lambda)
\end{equation}
where, as seen in Figure 1, $\theta_{ij}$ is the angular diameter
seen on Earth between the points on the star $i$ and $j$ and
$\lambda$ is the wavelength of the photons. $C_{ij}(d)$ is
proportional to the coincidence signal produced in the
photomultipliers for the photons coming from these two points on the
star, but in practice, the signal measured in the electronics is the
sum over all of the photon emission point pairs from the star. If
$N$ is defined as the number of photon source points making up a
star of brightness (intensity) $I$ and radius $r$, to get the
detected signal from the entire star, $C(d)$, $C_{ij}$ is summed
over all of the unique pairs of photon emission points
\begin{equation}
\label{hbt4}
 C(d)=\epsilon\sum^N_{i>
j}C_{ij}(d)=\epsilon\sum^N_{i>j}\cos(2\pi\theta_{ij}d/\lambda)
\end{equation}
where $\epsilon$ is a conversion factor to get the detector signal.
If one looks at cases where the arguments of the $cos$ terms in Eq.
(\ref{hbt4}) are small such that the $cos$ terms are not far from
unity, e.g. for small baselines, Eq. (\ref{hbt4}) can approximately
be expressed as
\begin{equation}
\label{hbt5}
C(d)\simeq\epsilon\frac{N^2}{2}\cos(2\pi\theta_1d/\lambda)
\end{equation}
where $\theta_1$ is the average of the $\theta_{ij}$.

One is now in the position to measure the average angular size of a
star by measuring $C(d)$ as a function of $d$ and fitting these data
with Eq. (\ref{hbt5}) to extract $\theta_1$. Figure 2 shows
measurements of $C(d)$ for Sirius by Hanbury- Brown and Twiss. Also
plotted are fits to the data using Eq. (\ref{hbt5}) (taking
$\lambda=550$ nm) and a full calculation including integrations over
the photon wavelength spectrum and photon source distribution and
detection efficiency effects by Hanbury-Brown and Twiss.\cite{hbt2}
As seen, $\theta_1$ extracted from the fit to the data by Eq.
(\ref{hbt5}) agrees with the angular diameter extracted from the fit
by the full calculation within a factor of two.

\begin{figure}[th]
\centerline{\psfig{file=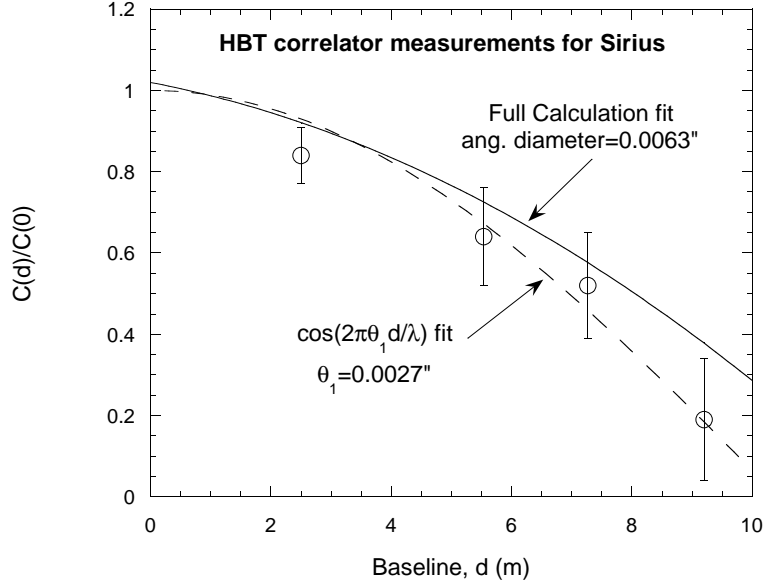,width=10cm}} \vspace*{8pt}
\caption{Measurement of $C(d)/C(0)$ vs. d for Sirius by
Hanbury-Brown and Twiss. Also shown are fits to the measurements
using Equation 4 and a full calculation.}
\end{figure}

Another example of applying the HBT technique to measure angular
size is for the binary star Alpha Centauri, located in the
constellation Centauris seen in the southern hemisphere. A schematic
diagram of the geometry of the measurement is shown in Figure 3. For
the orientation of the two stars shown, there are two angular size
scales that enter this measurement: the angular sizes of the
individual stars (assumed to be identical), $\theta_1$, and the
angular size between the two stars, $\theta_2$. What angular size
will the HBT technique measure? To answer this question, calculate
$C_{binary}(d)$ using Eq. (\ref{hbt4}) assuming that the binary star
has the total number of photon source points $N$ distributed equally
between the two stars and setting $\epsilon=1$ for simplicity,
giving
\begin{equation}
\label{hbt6}
 C_{binary}(d)=
2\sum^{N/2}_{i>j}\cos(2\pi\theta_{ij}d/\lambda)+
\sum^{N/2}_{i(A),j(B)}\cos(2\pi\theta_{ij}d/\lambda)
\end{equation}
where the first sum accounts for the two identical sums over the
individual stars in the binary, and the second sum accounts for the
sum over photon source pairs between the two stars, i.e. the index
$i$ associated with star A and $j$ with the star B. One can now use
the same approximation as before in going from Eq. (\ref{hbt4}) to
Eq. (\ref{hbt5}), i.e. the argument of the cosine being small, to
express the first sum in Eq. (\ref{hbt6}) in terms of $\theta_1$.
This will not be the case for the second sum, since for Alpha
Centauri the separation between the stars, $L$, is much larger than
the radius of the individual stars, $r$ ($L\simeq11$ A.U. and
$r\sim$ solar radius), resulting in $\theta_2>>\theta_1$. Thus, the
$cos$ terms in the second sum will oscillate wildly, resulting in
that sum vanishing, and Eq. (\ref{hbt6}) becomes
\begin{equation}
\label{alphac}
C_{binary}(d)=\frac{N^2}{4}\cos(2\pi\theta_1d/\lambda).
\end{equation}
Comparing Eq. (\ref{alphac}) with Eq. (\ref{hbt5}), it is seen that
the binary star gives half the detected HBT signal expected from a
star with $N$ photon source points, and its angular size is
characterized by the size of the individual stars making up the
binary. When Hanbury-Brown and Twiss performed such a measurement on
Alpha Centauri, this is indeed what they observed.\cite{cris2}

\begin{figure}[th]
\centerline{\psfig{file=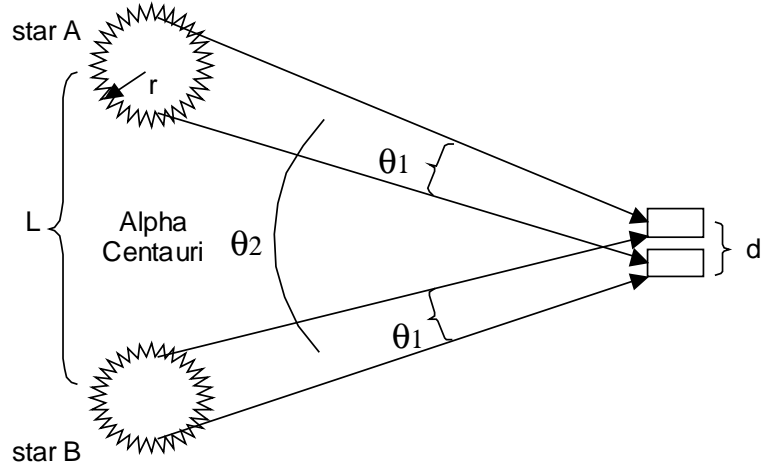,width=10cm}} \vspace*{8pt}
\caption{Schematic diagram of the geometry for the measurement of
the angular size of the binary star, Alpha Centauri.}
\end{figure}

This example of the use of this technique on a binary star shows that the
particular boson source distribution of the system measured has an important
impact on how results from HBT are interpreted. It will be shown later that
similar effects are seen when applying this technique to two-pion measurements
of RHI.

\section{Applying HBT to Relativistic Heavy-Ion Collisions}
Relativistic heavy-ion collisions are similar to stars in that they
emit bosons from a finite-sized region (i.e. the interaction
region). They differ from stars in that 1) the bosons are
predominantly pions, 2) other bosons such as kaons are emitted, 3)
the size scale is much smaller ($\sim10^{-15}$ m), and 4) the
lifetime of the boson-emitting source is short ($\sim10^{-23}$ s),
introducing time as an important variable. Nevertheless, one could
imagine applying HBT to RHI to directly measure the size, and
perhaps the lifetime, of these collisions and thus get information
on their space-time dynamics. The first measurement of the
pion-emitting source using two-pion interferometry was carried out
by G. Goldhaber, S. Goldhaber, W. Lee, and A. Pais in 1960 with
$\sqrt{s}=2.1$ GeV proton+antiproton collisions from the Bevatron
(they referred to this method as Bose-Einstein correlations and
others have called this the GGLP effect).\cite{gglp} In this section
a qualitative derivation is given for two-pion Bose-Einstein/HBT
interferometry and a few practical considerations for carrying out
experimental RHI HBT studies are given.

\subsection{Qualitative derivation of two-pion HBT in RHI}
A qualitative derivation of two-pion HBT interferometry is given
below. More formal derivations can be found in the
literature.\cite{gyu,heinz1999} As will be seen, its derivation will
closely parallel that given earlier for two-photon HBT. A schematic
diagram of the geometry of a two-pion interferometric measurement is
given in Figure 4. The interaction region in a RHI collision is
shown emitting two pions from points $\mathbf{x}_i$ and
$\mathbf{x}_j$, which are detected with wavevectors $\mathbf{k}_i$
and $\mathbf{k}_j$ in two ``pion detectors'' located at positions
$\mathbf{a}_1$ and $\mathbf{a}_2$, respectively, in the experimental
hall.

\begin{figure}[th]
\centerline{\psfig{file=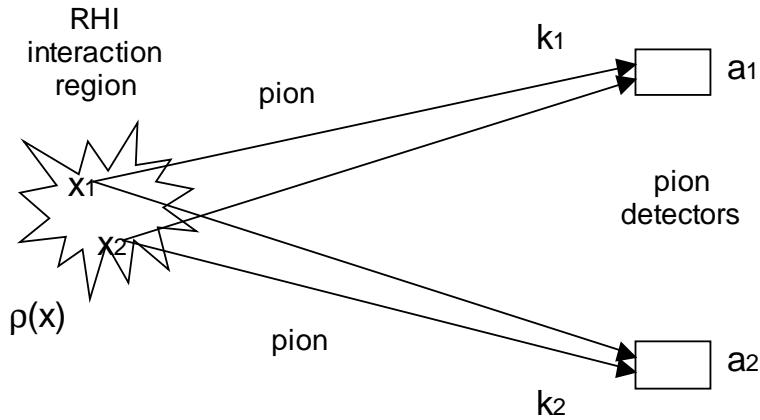,width=10cm}} \vspace*{8pt}
\caption{Schematic diagram of a two-pion interferometric measurement
in which the pions are produced in a RHI collision.}
\end{figure}

Assuming that pions are emitted incoherently from the interaction region
at each position and the pions are emitted as plane waves and taking a time-independent picture,
the wavefunction to detect in coincidence the two pions with
these wavevectors in the two detectors,
$\Psi(\mathbf{k}_i,\mathbf{k}_j;\mathbf{x}_i,\mathbf{x}_j)$, is
\begin{eqnarray}
\label{2pi1}
\Psi(\mathbf{k}_i,\mathbf{k}_j;\mathbf{x}_i,\mathbf{x}_j)=
b[\exp(i\mathbf{k}_i\cdot(\mathbf{a}_1-\mathbf{x}_i))
\exp(i\mathbf{k}_j\cdot(\mathbf{a}_2-\mathbf{x}_j)) \nonumber \\ +
\exp(i\mathbf{k}_i\cdot(\mathbf{a}_1-\mathbf{x}_j))
\exp(i\mathbf{k}_j\cdot(\mathbf{a}_2-\mathbf{x}_i))]
\end{eqnarray}
where the second term arises due to the path ambiguity of detecting
bosons (as shown in Figure 4), and where $b$ is a normalization
constant. This expression is similar to Eq. (\ref{hbt1}) above used
for photon pairs from a star, but not identical. The difference is
that indices are switched in the second term since we assume the
pion wavevector (momentum) is now associated with a particular
detector, rather than with an emission point as in the case of a
photon from the star. The ambiguity in this case is in which
emission point the pion originates from, not which detector records
a photon. The probability to detect the two pions, $P_{ij}$, is just
the square of $\Psi$, i.e. $\Psi^*\Psi$, and thus given by
\begin{equation}
\label{prob2pi}
P_{ij}(\Delta\mathbf{k},\mathbf{r})=
b^2[1+\cos(\Delta\mathbf{k}\cdot\mathbf{r})]
\end{equation}
where $\Delta\mathbf{k}=\mathbf{k}_i-\mathbf{k}_j$ and
$\mathbf{r}=\mathbf{x}_i- \mathbf{x}_j$. The separation between the
two emission points, $r$, is related to the size of the interaction
region or ``pion source.'' It is seen in Eq. (\ref{prob2pi}) that a)
the $cos$ term in $P_{ij}$ is a result of the path ambiguity term in
Eq. (\ref{2pi1}), and b) $P_{ij}$ does not depend on the detector
positions but only on the differences between the pion momenta and
the emission positions. Note that the time variable can also be
introduced into this formalism by considering $\mathbf{k}$ and
$\mathbf{x}$ as four vectors.

As in the discussion of HBT with stars in the last section, the
interaction region can emit pions from an extended region of points
in space which the detectors sum over. Defining the pion source
distribution function as a continuous function of the position of a
source point $\mathbf{x}$, $\rho(\mathbf{x})$, one can integrate the
interaction region over all pairs of source points using Eq.
(\ref{prob2pi}),

\begin{equation}
\label{corr2pi}C(\Delta\mathbf{k})=\int
P_{ij}(\Delta\mathbf{k},\mathbf{r})\rho(\mathbf{x_i})
\rho(\mathbf{x_j})d\mathbf{x_i}d\mathbf{x_j},
\end{equation}

where $C(\Delta\mathbf{k})$ is called the correlation function.
Assuming for simplicity that the pion source is distributed as a
gaussian of width $R$, i.e. $\rho(r)\sim\exp(-r^2/2R^2)$, and
substituting into Eq.(\ref{corr2pi}) the result is,

\begin{equation}
\label{corr2pig}C(\Delta\mathbf{k})\sim
1+\exp[-(\Delta\mathbf{k})^2R^2].
\end{equation}

Figure 5. shows a two-pion HBT measurement in central RHIC Au+Au
collisions from the STAR experiment, which detects charged particles
in a large-acceptance Time Projection Chamber immersed in a magnetic
field.\cite{Adler:2001b} The one-dimensional two-pion correlation
function (extracted from the two-pion coincidence count rate, see
later) evaluated in the invariant frame of the pion source is
plotted versus $\Delta k$. Only the data points lower than
$\hbar\Delta k < 50$ MeV/c are plotted. A version of Eq.
(\ref{prob2pi}) assuming $\Delta\mathbf{k}\perp\mathbf{r}$ including
an arbitrary parameter $\lambda$ is fitted to the data to extract
$r$ and $\lambda$ (dashed line). A version of Eq. (\ref{corr2pig})
with the same parameters is also fitted (solid line). As seen, the
gaussian function fits the data better, but both functions give
about the same fitted values of $r\sim 8$ fm and $\lambda\sim
0.5-0.6$.

\begin{figure}[t]
\centerline{\psfig{file=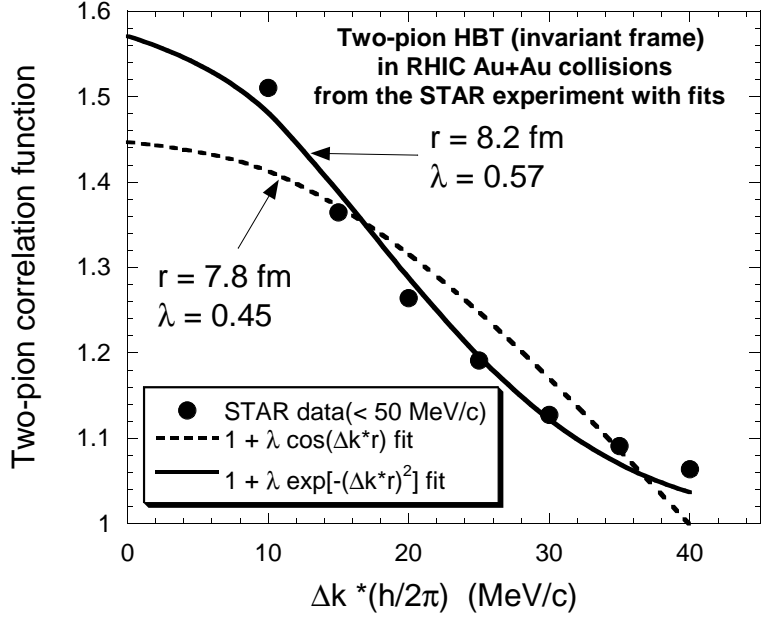,width=10cm}} \vspace*{8pt}
\caption{Two-pion HBT in RHIC Au+Au collisions from the STAR
experiment with fits. Only points for $\hbar\Delta k < 50$ MeV/c are
plotted.}
\end{figure}

What is the interpretation of these results for $r$ and $\lambda$?
Naively, one would expect the size of the Au nucleus,
$1.2A^{1/3}\sim 7$ fm, to set the scale for the size of the pion
emitting source in these collisions, and this is seen to be the case
for $r$. For $\lambda$ one might expect this to be unity for the
idealized derivation of Eq. (\ref{prob2pi}), yet the value extracted
in this case is significantly smaller. This is suggestive of the
reduced HBT signal seen in binary stars discussed above, for which
boson sources of distinctly different sizes are intermixed.

How to interpret the parameters from such fits to data in RHI HBT
analyses will be an important theme for the rest of this paper. As
will be shown, models of the interaction will be used as tools to
interpret experimental HBT results.

\subsection{Measuring the Correlation Function in experiments}
As shown in Figure 4., in carrying out an experimental HBT study one
measures the two-boson coincident count rate along with the
one-boson count rates for reference. The experimental two-boson
correlation function for bosons binned in momenta $\mathbf{k_1}$ and
$\mathbf{k_2}$, $C(\mathbf{k_1},\mathbf{k_2})$, is constructed from
the coincident countrate, $N_2(\mathbf{k_1},\mathbf{k_2})$ and
one-boson countrate, $N_1(\mathbf{k})$, as
\begin{equation}
\label{correxp}
C(\mathbf{k_1},\mathbf{k_2})=\alpha(\mathbf{k_1},\mathbf{k_2})\frac{N_2(\mathbf{k_1},
\mathbf{k_2})}{N_1(\mathbf{k_1})N_1(\mathbf{k_2})},
\end{equation}
where $\alpha(\mathbf{k_1},\mathbf{k_2})$ is a correction factor for
non-HBT effects which may be present in the experiment.
Typically,the largest contributions to
$\alpha(\mathbf{k_1},\mathbf{k_2})$ occur in the correction for the
boson detection efficiency and in correcting for final-state
boson-boson Coulomb repulsion.\cite{lisa2005} It is usually
convenient to express the six-dimensional
$C(\mathbf{k_1},\mathbf{k_2})$ in terms of the four-vector momentum
difference, $Q=|k_1-k_2|$ by summing Eq.(\ref{correxp}) over
momentum difference,
\begin{eqnarray}
\label{correxp2} C(Q)=\sum_{\mathbf{k_1},\mathbf{k_2}(Q)}
\alpha(\mathbf{k_1},\mathbf{k_2})\frac{N_2(\mathbf{k_1},
\mathbf{k_2})}{N_1(\mathbf{k_1})N_1(\mathbf{k_2})}=
\epsilon(Q)\frac{A(Q)}{B(Q)},
\end{eqnarray}
where $A(Q)$ represents the ``real'' coincident two-boson countrate,
$B(Q)$ the ``background'' two-boson countrate composed of products
of the one-boson countrates and $\epsilon(Q)$ the correction factor,
all expressed in $Q$.~\cite{heinz1999} In practice, $B(Q)$ is the
mixed event distribution, which is computed by taking single bosons
from separate events.\cite{lisa2005}

\subsection{Parameterizing the Correlation Function: extracting boson source
parameters} The experimental two-boson correlation function is
formed using Eq.(\ref{correxp2}) and a gaussian model for the boson
source distribution is normally fitted to it to extract radius
parameters. A standard parameterization of C(Q) is obtained by
assuming a gaussian space-time distribution of freeze-out points,
$\rho(r,t)$, in terms of the variables $r_{Tside}$, which points in
the direction of the sum of the two boson momenta in the transverse
plane, $r_{Tout}$, which points perpendicular to $r_{Tside}$ in the
transverse plane, the longitudinal variable along the beam
direction, z, and time, t:
\begin{eqnarray}
\label{e8} \rho(r,t) = F \exp( -
\frac{{r_{Tside}}^2}{2{R'_{Tside}}^2} -
\frac{{r_{Tout}}^2}{2{R'_{Tout}}^2} - \frac{z^2}{2{R'_{Long}}^2} -
\frac{t^2}{2\tau^2})
\end{eqnarray}
where $R'_{Tside}$ is a transverse sideward radius parameter,
$R'_{Tout}$ is a transverse outward radius parameter, $R'_{Long}$ is
a longitudinal radius parameter, $\tau$ is a lifetime parameter, and
F is a normalization constant.\cite{bek} From this distribution
function the following parameterization of C(Q) can be obtained
:\cite{pra2}
\begin{eqnarray}
\label{e9}
\lefteqn{C(Q_{Tside},Q_{Tout},Q_{Long}) = } \nonumber \\
& & G[ 1 + \lambda \exp( - Q_{Tside}^{2}R_{Tside}^{2} -
Q_{Tout}^{2}R_{Tout}^{2} - Q_{Long}^{2}R_{Long}^{2}) ]
\end{eqnarray}
where Q has been broken up into the two transverse and one
longitudinal components, G is a normalization constant, and
$\lambda$ is the usual empirical parameter added to help in the
fitting of Eq. (\ref{e9}) to the actual correlation function
($\lambda = 1$ in the ideal case). The radius parameters in Eq.
(\ref{e9}) are related to those in Eq.(\ref{e8}) in, for example,
the LCMS frame (longitudinally comoving system in which the
longitudinal boson pair momentum vanishes) as follows:
\begin{eqnarray}
\label{e10}
R_{Tside}^{2}& = &{R'_{Tside}}^{2} \nonumber \\
R_{Tout}^{2}& = &{R'_{Tout}}^{2} + \beta_T^{2}\tau^{2} \\
R_{Long}^{2}& = &{R'_{Long}}^{2} \nonumber
\end{eqnarray}
where  $\beta_T$ is the transverse velocity of the boson pair. In
this parameterization as seen from Eq. (\ref{e10}) $R_{Tout}$
contains information about both the transverse size and the lifetime
of the source. Note that Eq.(\ref{e9}) follows from Eq.(\ref{e8})
under the assumption of a geometrically static boson source. In a
realistic heavy-ion collision the source will not be static and may
have position-momentum correlations and other effects that could
make the source parameters defined above depend on the boson pair
momentum . In the present application, Eq. (\ref{e9}) is fitted to
the experimental correlation function to extract the radius
parameters $R_{Tside}$, $R_{Tout}$, and $R_{Long}$ and $\lambda$.
Figure 6 shows experimental $\pi^--\pi^-$ correlation functions with
and without final-state Coulomb corrections included from the
Relativistic Heavy Ion Collider (RHIC) STAR
experiment\cite{Adler:2001b} for $\sqrt{s}=130$ GeV/nucleon Au+Au
collisions along with gaussian fits.
\begin{figure}[t]
\centerline{\psfig{file=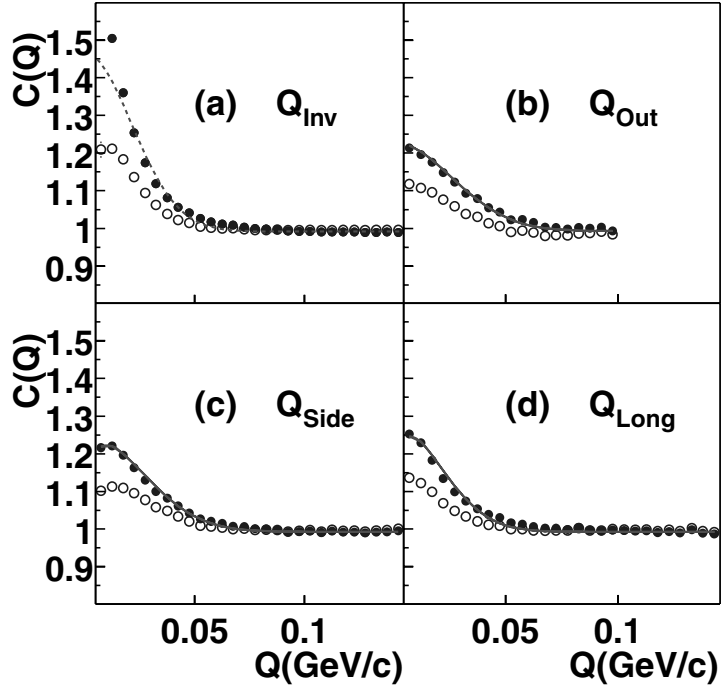,width=10cm}} \vspace*{8pt}
\caption{$\pi^--\pi^-$ HBT in RHIC Au+Au collisions from the STAR
experiment with gaussian fits. The Coulomb-corrected data are shown
as filled circles and the uncorrected data as open circles (from
Ref.[10]).}
\end{figure}

\section{Comparing hadronic scattering models with two-boson HBT
experiments from RHI collisions} Relativistic heavy-ion collisions
provide a means of creating matter in a hot and dense state which
might shed light on the behavior of matter under these extreme
conditions, such as the possibility of producing a phase transition
to quark matter.\cite{mul} It is generally agreed that the most
extreme conditions exist in the initial state of the heavy-ion
collision, roughly defined as occurring just after the projectile
and target nuclei pass through each other. Eventually the
interaction region hadronizes into a large number of mesons and
baryons (with non-hadronic particles such as photons, electrons, and
muons also being produced) and then expands to its final state.
During the expansion stage the hadrons strongly scatter with each
other, this process being called rescattering. The final state of
the collision can be thought of as the state for which rescattering
ceases among all remaining (final) hadrons. It is often convenient
to define the ``freezeout'' point of a final hadron as the position,
time, energy, and momentum the particle had when it stopped
rescattering. Thus, one can define more precisely the final state of
the collision as the collection of the freezeout points of all of
the final hadrons in an 8-dimensional phase space (4 dimensions for
space-time and 4 dimensions for momentum-energy). The term freezeout
will be used to represent the final state of the collision defined
in this way. In principle, the properties of the collision at
freezeout are directly accessible by measurement. In practice, one
directly measures the freezeout momenta and energies of the
particles using, for example, magnetic spectrometers \cite{bek} and
indirectly measures the freezeout space-time from the freezeout
momenta and energies using the method of two-boson HBT
interferometry.

However, the main motivation to study relativistic heavy-ion
collisions is to obtain information about the initial, extreme state
of the collision. Some directly measurable non- hadronic probes such
as direct photons, and electron and muon pairs have been predicted
to be sensitive to certain features of the initial state.\cite{sat}
It is equally important to study the final hadrons from the
collision since they should contain information about the bulk
properties of the initial state, i.e. the temperature and energy
density achieved in the collision. The difficulty in using the
hadrons to extract this information is that the rescattering process
masks the initial space-time and momentum-energy information by
random scattering and thus there is no simple connection between the
freezeout information obtained in experiments and the initial state.

In this paper, the method employed to approach this problem is to
use hadronic scattering calculations to disentangle the rescattering
effects from the hadronization process.  The strategy will be to
take a simple model for hadronization and propagate the initial
hadrons via rescattering to freezeout, adjusting the parameters of
this model to see if a parameter set can be found where the
freezeout observables from the calculation agree with those measured
in experiments. Within the context of such a model, this parameter
set thus describes the state of the collision before rescattering,
putting one a step closer in time to the initial state. The
advantage of using a simple hadronization model is that the number
of parameters to be adjusted is minimized, increasing the chances
that the extracted parameter set is unique. The disadvantage is that
some physics of the hadronization will be left out, so the physical
interpretation of these parameters may be complicated. Descriptions
of two such hadronic scattering models are presented below. Results
of these models are then compared with RHI HBT experiments.

\subsection{Intranuclear Cascade Model for $\sqrt{s}<$ 1
GeV/nucleon: INCM} Intranuclear cascade models (INCM) have been used
extensively to understand general features of RHI collisions for
bombarding energies of $\sqrt{s}<$ 1 GeV/nucleon.\cite{cascade1} The
basic assumption in these models is that a RHI collision can be
viewed as a superposition of nucleon-nucleon interactions whose
trajectories between interactions are described classically while
the interactions are determined by experimental scattering cross
sections. The CASCADE code by Cugnon\cite{cascade2} is a version of
a INCM. In this version, $\Delta$ isobars are included which serve
as the mechanism through which pions are produced and rescattered
throughout the duration of the collision. The CASCADE code is
isospin averaged such that there is only one type of pion, nucleon,
and $\Delta$ in the calculation. A pion is defined as freezing out
from the calculation when it no longer scatters and the time and
position of its parent $\Delta$, along with the vector momentum of
the pion, are recorded.

Two-pion HBT predictions are made with this recorded information by
weighting pairs of pions with bose symmetrization using a
four-vector version of Eq. (\ref{prob2pi}) (so that time effects are
included) and then binning the weighted pion pairs similar to what
is done in experiments using Eq. (\ref{correxp2})to form the
two-pion correlation function.\cite{humanic1986} Predictions for
pion source parameters are then obtained by fitting a gaussian model
to this Monte-Carlo generated correlation function. The particular
pion source model used in this study is similar to that represented
by Eq. (\ref{e8}) but with a spherical gaussian source, i.e.
$\rho(r,t) \sim \exp(-\frac{r^2}{R^2}-\frac{t^2}{\tau^2})$, giving
\begin{equation}
\label{corr2tau} C(Q,Q_0)=1+\lambda
\exp(-\frac{Q^2R^2}{2}-\frac{{Q_0}^2\tau^2}{2})
\end{equation}
where the $R$ and $\tau$ parameters differ by a factor of $\sqrt{2}$
from the similar quantities defined in Eq. (\ref{e8}), the $\lambda$
parameter has the same meaning as before,
$Q=|\mathbf{k_1}-\mathbf{k_2}|$, and $Q_0={\mid{E_1-E_2}\mid}$,
where $E$ is the total relativistic energy of a pion. Figure 7 shows
a two-pion correlation function projected onto the $Q$ variable
generated from this procedure using CASCADE for the reaction 1.5
GeV/nucleon (lab frame) $Ar^{40}+Ar^{40}$ with impact parameter
$b=2$ fm and minimum pion momentum $k_{min}=50$ MeV/c. The projected
fit of Eq. (\ref{corr2tau}) is also shown. The source parameters
extracted from the model in this case are $R=3.6\pm0.1$ fm,
$\tau=3.2\pm0.5$ fm/c, and $\lambda=0.94\pm0.05$. These parameters
are shown in the larger context of their dependence on $k_{min}$ and
$b$ in Figure 8. As seen, for increasing $k_{min}$ with fixed $b$,
$R$ and $\tau$ are seen to decrease somewhat whereas $\lambda$ stays
constant at around unity. For increasing $b$ and fixed $k_{min}$,
all three parameters are seen to decrease.

\begin{figure}[t]
\centerline{\psfig{file=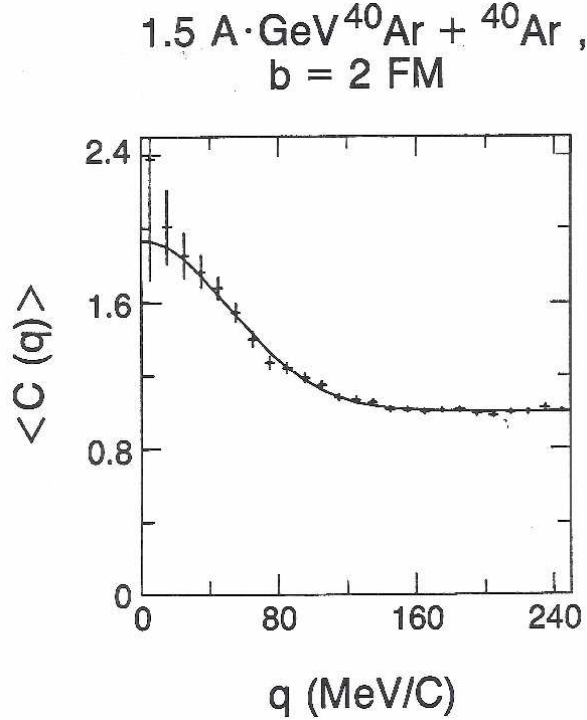,width=8cm}} \vspace*{8pt}
\caption{Projected two-pion correlation function from CASCADE for
the reaction 1.5 GeV/nucleon $Ar^{40}+Ar^{40}$, $b=2$ fm,
$k_{min}=50$ MeV/c (From Ref.[17]).}
\end{figure}

\begin{figure}[t]
\centerline{\psfig{file=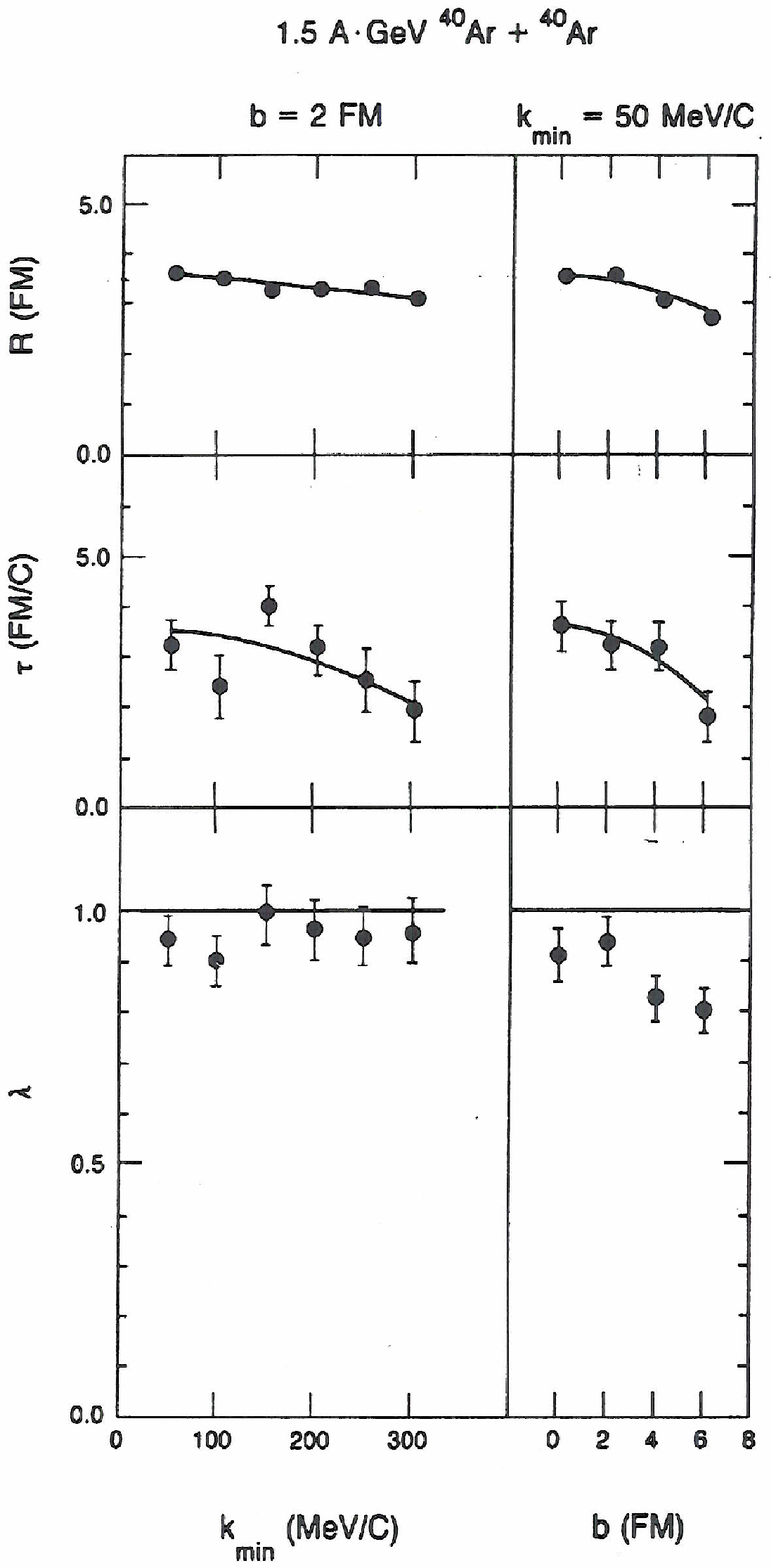,width=6cm}} \vspace*{8pt}
\caption{Dependence of pion source parameters on $k_{min}$ and $b$
for the reaction 1.5 GeV/nucleon $Ar^{40}+Ar^{40}$. The lines are
drawn to guide the eye (From Ref.[17]).}
\end{figure}

The mechanisms for these dependencies in the model are seen in
Figures 9 and 10. Figure 9 shows that the freezeout time
distribution for larger $k_{min}$ is shifted backward in time
compared with smaller $k_{min}$. Since the pion source from the
model is found to be smaller at earlier times, this explains the
smaller $R$ and $\tau$ for larger $k_{min}$: higher pion momentum
probes earlier stages of the collision. Figure 10 shows the effect
of impact parameter on the size and shape of the pion source. For
small $b$ the source projected on the $x-z$ plane is spherically
symmetric and large, while for larger $b$ the source is smaller and
extended along the $z$ axis. Thus one expects the decrease in $R$
and $\tau$ for larger $b$, and sees that the decrease of $\lambda$
is due to the source no longer looking like a perfect gaussian.

\begin{figure}[t]
\centerline{\psfig{file=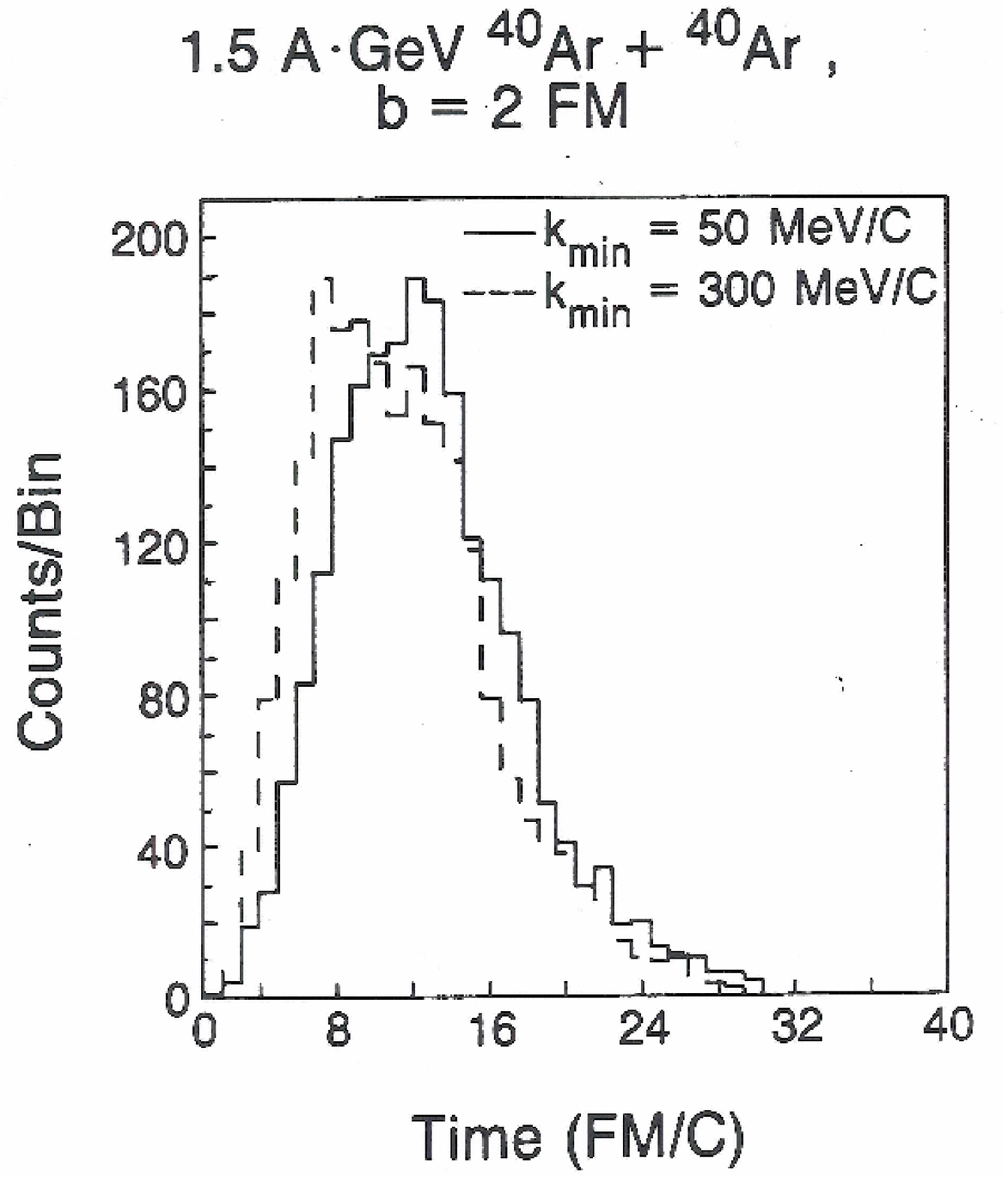,width=8cm}} \vspace*{8pt}
\caption{Comparison between final pion freezeout time distributions
for $k_{min}=50$ and $300$ MeV/c, for the reaction 1.5 GeV/nucleon
$Ar^{40}+Ar^{40}$ (From Ref.[17]).}
\end{figure}

\begin{figure}[t]
\centerline{\psfig{file=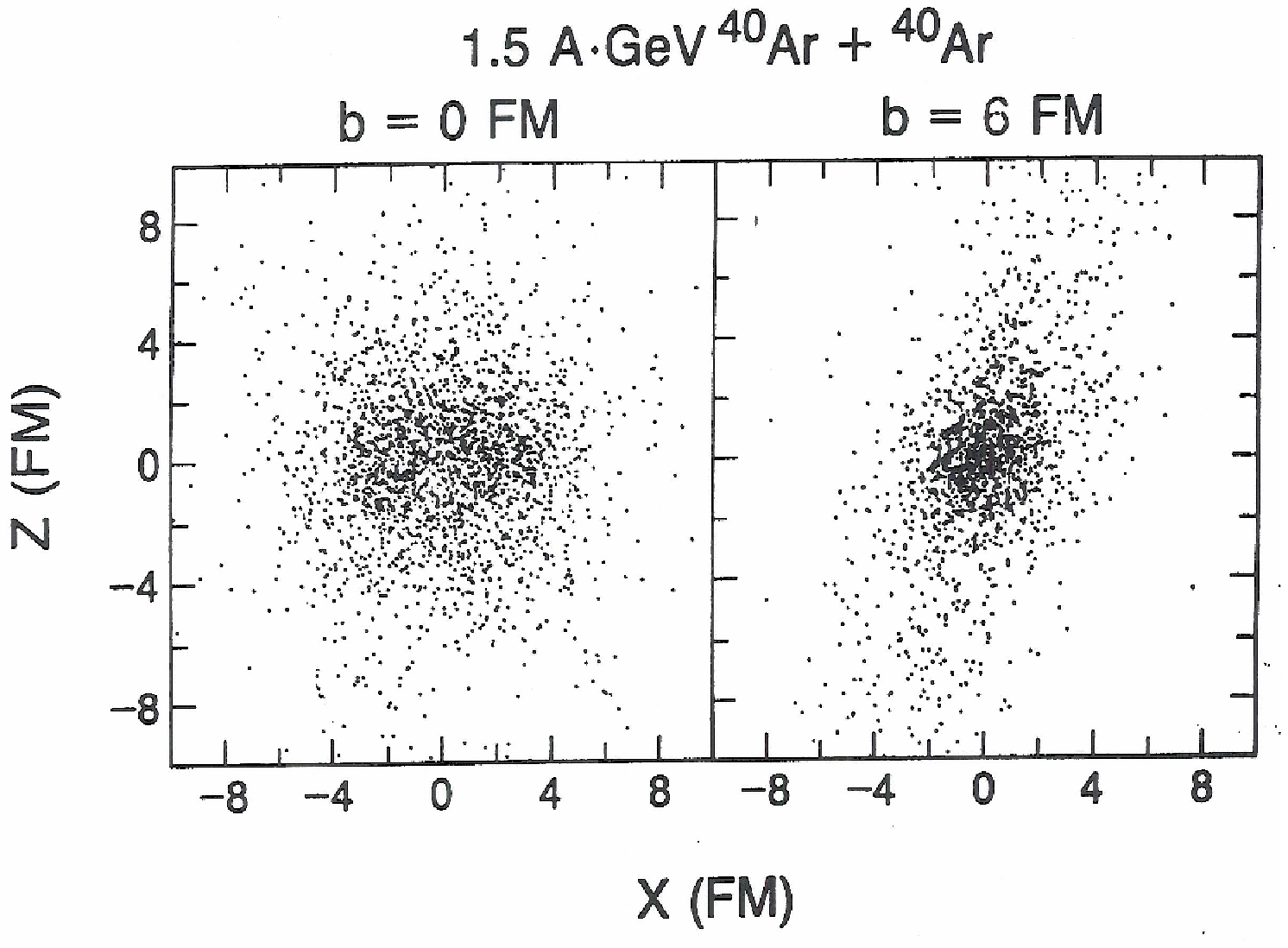,width=10cm}} \vspace*{8pt}
\caption{Time-integrated spacial pion source distribution projected
onto the reaction plane for $b=0$ and $6$ fm, for the reaction 1.5
GeV/nucleon $Ar^{40}+Ar^{40}$ (From Ref.[17]).}
\end{figure}

\subsection{Comparing the INCM with Bevalac experiments}
In the last section HBT predictions based on the CASCADE INCM were
presented and analyzed. How seriously to take these results depends
on whether or not this model gives predictions which agree with
experiment. Figure 11 shows a comparison of pion source parameters
extracted from the model with those from several LBL Bevalac HBT
experiments: (a) 1.8 GeV/nucleon $Ar^{40}+KCl$ and
$Ne^{20}+NaF$,\cite{bevalac1} (b) 1.5 GeV/nucleon
$Ar^{40}+KCl$,\cite{bevalac2,bevalac3} and (c) 1.2 GeV/nucleon
$Ar^{40}+KCl$.\cite{bevalac3} The experiments used two different
types of detectors to measure the pion momenta, a narrow acceptance
magnetic spectrometer\cite{bevalac1} and a large acceptance streamer
chamber.\cite{bevalac2,bevalac3} For the spectrometer, both
$\pi^+\pi^+$ and $\pi^-\pi^-$ pairs were used. In order to simulate
the pion acceptances used in the two experiments, pions used in the
model predictions were selected to be only those which fell in the
experimental acceptances.

\begin{figure}[t]
\centerline{\psfig{file=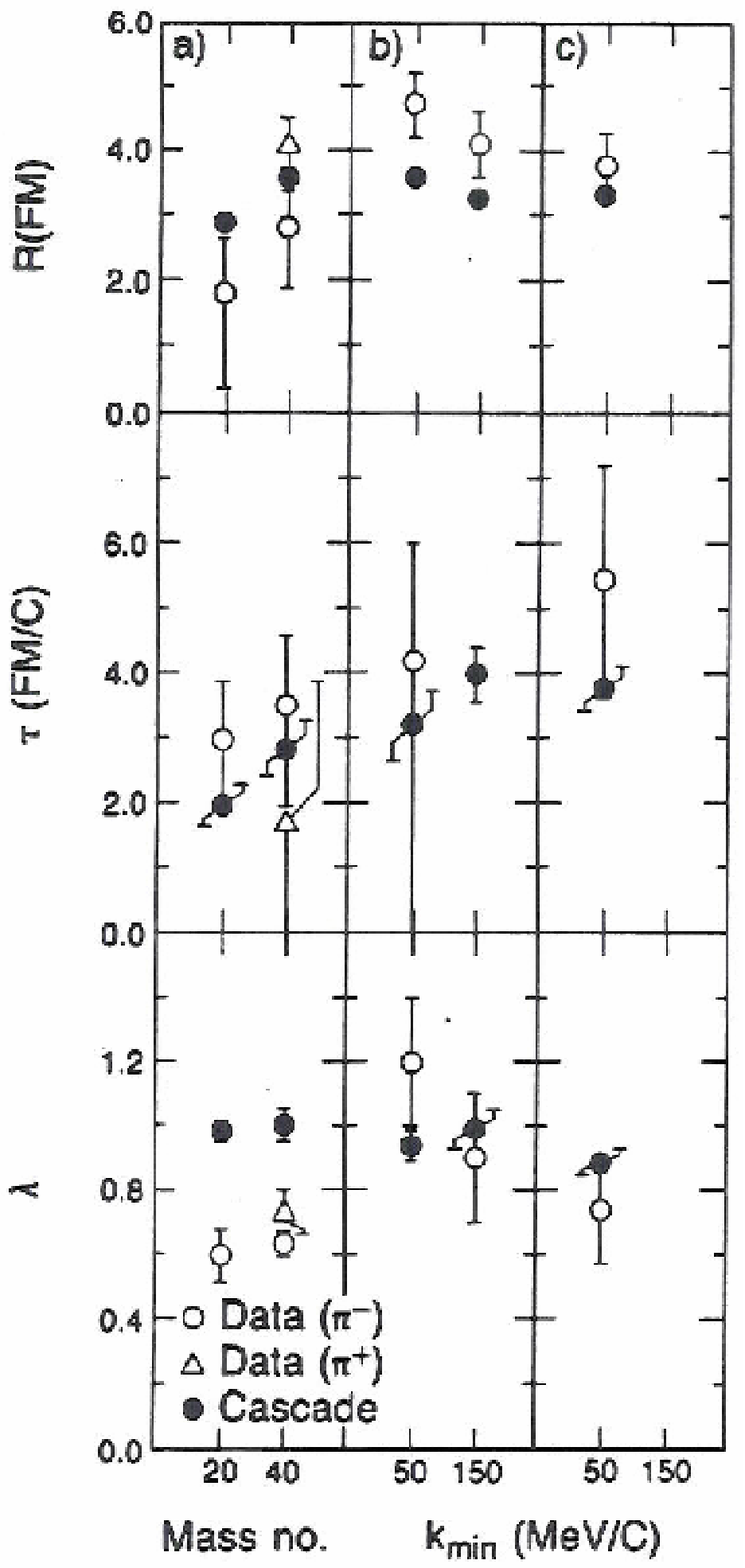,width=6cm}} \vspace*{8pt}
\caption{Comparison between pion source parameters from CASCADE
predictions and pion interferometry measurements (a) 1.8 GeV/nucleon
$Ar^{40}+KCl$ and $Ne^{20}+NaF$, (b) 1.5 GeV/nucleon $Ar^{40}+KCl$,
and (c) 1.2 GeV/nucleon $Ar^{40}+KCl$ (From Ref.[17]).}
\end{figure}

As seen in Figure 11, the model predictions, labeled as ``Cascade'',
mostly agree with the experimental source parameters within
$1\sigma$ of the experimental error bars. The largest disagreement
is seen for the spectrometer $\lambda$ parameters, the predictions
being close to unity and the measurements being in the $0.5-0.7$
range, whereas from the streamer chamber experiment $\lambda$ is
measured to be close to unity as with the predictions.

From the overall good agreement between model and experiments, one
can conclude that the method of symmetrizing the pions produced in
the intranuclear cascade hadronic scattering model to make HBT
predictions can be a valuable tool in understanding what HBT is
measuring at the ``microscopic level'' in these lower energy RHI
collisions. This motivates one to apply this same method to hadronic
scattering models valid for higher energy RHI collisions. This is
done in the next section.

\subsection{Hadronic Rescattering Model for $\sqrt{s}>$ 1 GeV/nucleon: HRM}
For higher energy RHI collisions, i.e. $\sqrt{s}>$ 1 GeV/nucleon, a
different hadronic scattering approach is taken to simulate the
collision as the basis for HBT predictions: a rescattering
calculation is used to disentangle the rescattering effects from the
hadronization process. As mentioned earlier, the strategy will be to
take a simple model for hadronization and propagate the initial
hadrons via rescattering to freeze out, adjusting the parameters of
this model to see if a parameter set can be found where the
freeze-out observables from the calculation agree with those
measured in experiments. Descriptions of both the hadronization
model and the rescattering calculation used are presented below.
Results of applying this approach for the BNL Alternating Gradient
Synchrotron (AGS), CERN Super Proton Synchrotron (SPS), BNL
Relativistic Heavy Ion Collider (RHIC), and CERN Large Hadron
Collider (LHC) energy collisions are then shown.

\subsubsection{HRM description}
A brief description of the hadron rescattering model (HRM)
calculational method is given below. A more detailed description is
given elsewhere.\cite{humanic1998,humanic2003} Rescattering is
simulated with a semi-classical Monte Carlo calculation which
assumes strong binary collisions between hadrons. The Monte Carlo
calculation is carried out in three stages: 1) initialization and
hadronization, 2) rescattering and freeze out, and 3) calculation of
experimental observables. Relativistic kinematics is used
throughout.  All calculations are made to simulate either AGS, SPS,
RHIC, or LHC energy collisions in order to compare with present (or
future) experimental results

The hadronization model employs simple parameterizations to describe
the initial momenta and space-time of the hadrons similar to that
used by Herrmann and Bertsch.\cite{Herrmann:1995a} The initial
momenta are assumed to follow a thermal transverse (perpendicular to
the beam direction) momentum distribution for all particles,
\begin{equation}
\label{temp}
(1/{m_T})dN/d{m_T}=C{m_T}/[\exp{({m_T}/T)} \pm 1]
\end{equation}
where ${m_T}=\sqrt{{p_T}^2 + {m_0}^2}$ is the transverse mass, $p_T$
is the transverse momentum, $m_0$ is the particle rest mass, $C$ is
a normalization constant, and $T$ is the initial ``temperature'' of
the system, and a gaussian rapidity distribution for mesons,
\begin{equation}
dN/dy=D \exp{[-{(y-y_0)}^2/(2{\sigma_y}^2)]}
\end{equation}
where $y=0.5\ln{[(E+p_z)/(E-p_z)]}$ is the rapidity, $E$ is the
particle energy, $p_z$ is the longitudinal (along the beam
direction) momentum, $D$ is a normalization constant, $y_0$ is the
central rapidity value (mid-rapidity), and $\sigma_y$ is the
rapidity width. Two rapidity distributions for baryons have been
tried: 1) flat and then falling off near beam rapidity and 2) peaked
at central rapidity and falling off until beam rapidity. Both baryon
distributions give about the same results. The initial space-time of
the hadrons for $b=0$ fm (i.e. zero impact parameter or central
collisions) is parameterized as having cylindrical symmetry with
respect to the beam axis. The transverse particle density dependence
is assumed to be that of a projected uniform sphere of radius equal
to the projectile radius, $R$ ($R={r_0}A^{1/3}$, where ${r_0}=1.12$
fm and $A$ is the atomic mass number of the projectile). For $b>0$
(non-central collisions) the transverse particle density is that of
overlapping projected spheres whose centers are separated by a
distance b. The longitudinal particle hadronization position
($z_{had}$) and time ($t_{had}$) are determined by the relativistic
equations,\cite{Bjorken:1983a}
\begin{eqnarray}
\label{rel} z_{had}=\tau_{had}\sinh{y}\\
 t_{had}=\tau_{had}\cosh{y}\nonumber
\end{eqnarray}
where $y$ is the particle rapidity and $\tau_{had}$ is the
hadronization proper time. Thus, apart from particle multiplicities,
the hadronization model has three free parameters to extract from
experiment: $\sigma_y$, $T$ and $\tau_{had}$. The hadrons included
in the calculation are pions, kaons, nucleons and lambdas ($\pi$, K,
N, and $\Lambda$), and the $\rho$, $\omega$, $\eta$, ${\eta}'$,
$\phi$, $\Delta$, and $K^*$ resonances. For simplicity, the
calculation is isospin averaged (e.g. no distinction is made among a
$\pi^{+}$, $\pi^0$, and $\pi^{-}$). Resonances are present at
hadronization and also can be produced as a result of rescattering.
Initial resonance multiplicity fractions are taken from Herrmann and
Bertsch,\cite{Herrmann:1995a} who extracted results from the HELIOS
experiment.\cite{Goerlach:1992a} The initial resonance fractions
used in the present calculations are: $\eta/\pi=0.05$,
$\rho/\pi=0.1$, $\rho/\omega=3$, $\phi/(\rho+\omega)=0.12$,
${\eta}'/\eta=K^*/\omega=1$ and, for simplicity, $\Delta/N=0$. Note
that the AGS $Si+Au$ and SPS $S+S$ and $S+Pb$ calculations were done
with an earlier version of HRM and differ from the description above
in that 1) an initial cylinder was used instead of Eq. (\ref{rel})
of length $z=\pm1$ fm and 2) no resonances were
included.\cite{humanic1996,humanic1994}

The second stage in the calculation is rescattering which finishes
with the freeze out and decay of all particles. Starting from the
initial stage ($t=0$ fm/c), the positions of all particles are
allowed to evolve in time in small time steps ($dt=0.1$ fm/c)
according to their initial momenta. At each time step each particle
is checked to see a) if it decays, and b) if it is sufficiently
close to another particle to scatter with it. Isospin-averaged
s-wave and p-wave cross sections for meson scattering are obtained
from Prakash et al..\cite{Prakash:1993a} The calculation is carried
out to 100 fm/c, although most of the rescattering finishes by about
30 fm/c.

Calculations are carried out assuming initial parameter values and
particle multiplicities for each type of particle. In the last stage
of the calculation, the freeze-out and decay momenta and space-times
are used to produce observables such as pion, kaon, and nucleon
multiplicities and transverse momentum and rapidity distributions.
The values of the initial parameters of the calculation and
multiplicities are constrained to give observables which agree with
available measured hadronic observables. As a cross-check on this,
the total kinetic energy from the calculation is determined and
compared with the collision center of mass energy to see that it is
in reasonable agreement. Particle multiplicities were estimated from
charged hadron multiplicity measurements and models. The
hadronization model parameters used for various systems are shown in
Table \ref{tab:tab1}. It is interesting to note that it is desirable
the same value of $\tau_{had}$ for all three very disparate in
energy systems.

\begin{table}
\begin{center}
\caption{Hadronization model parameters used in the HRM for various
systems.}

    \label{tab:tab1}
\begin{tabular}{cccc}\hline
$System$ & $\tau_{had}$ (fm/c) & $T_{init}$ (MeV) & $\sigma_{yinit}$
\\\hline SPS Pb+Pb & 1                 & 270          & 1.2 \\
 RHIC Au+Au & 1                 & 300          & 2.4 \\
 LHC Pb+Pb & 1                 & 500          & 4.2 \\\hline
\end{tabular}
\end{center}
\end{table}

Figures 12-18 show some results from HRM to give a feeling for the
information one can obtain from this model.

Figures 12 and 13 show the effects of momentum cuts, both in
magnitude and direction, on the transverse and longitudinal pion
source dimensions, respectively. Figure 12 shows distributions of
pion freeze out positions for central collision SPS $S+Pb$
collisions projected onto the transverse plane. In (a) and (c) a low
momentum cut on the pions of $0<p_T<100$ MeV/c is made, while for
(b) and (d) a high pion momentum cut of $500<p_T<600$ MeV/c is made.
Also, in (c) and (d) only pions with azimuthal direction in the
range $165^o<\phi<195^o$ (as indicated by the arrows) are shown.
Comparing (a) and (b), one sees that higher momentum pions tend to
be concentrated at larger radius compared with the lower momentum
pions which tend to be uniformly distributed in a spherical volume.
As seen in (c) and (d), if on top of these momentum cuts on
magnitude cuts on direction are also made, the geometry of the low
momentum pion source is not significantly changed (aside from fewer
pions satisfying both cuts), whereas the higher momentum pions are
seen to be more directional and only pions in the vicinity of the
angular cut are present, making the pions source look smaller in
size. Figure 13 shows a similar effect along the ``light cone'' of
cutting on the direction of pions normal to the longitudinal
direction, seen in the lower plot of the figure. Since HBT tends to
pick out pion pairs with small momentum difference, the
distributions in Figure 12(c) and (d) and Figure 13 give a
qualitative indication of how HBT will view these cases.

\begin{figure}[th]
\centerline{\psfig{file=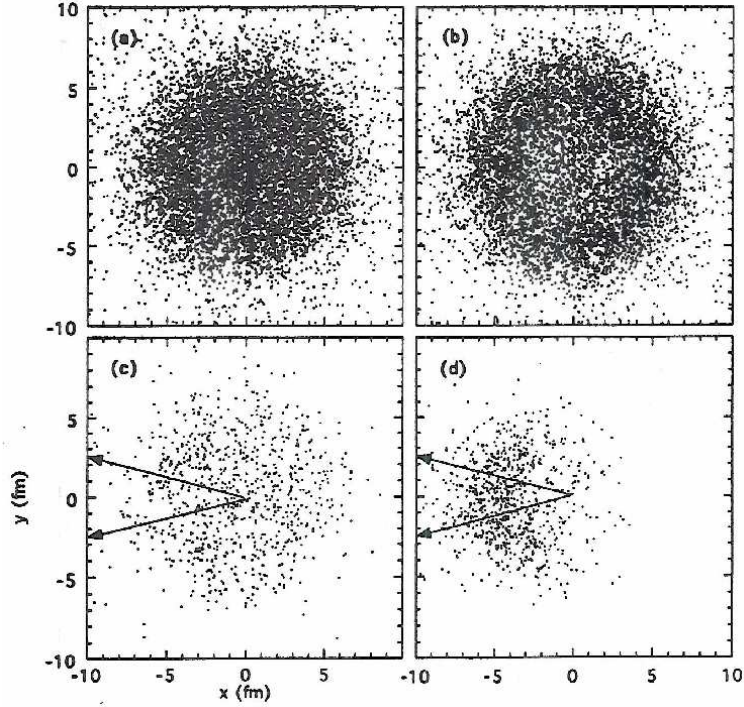,width=10cm}} \vspace*{8pt}
\caption{Distribution of pion freeze out positions for SPS $S+Pb$
collisions projected onto the transverse plane. (a) and (c):
$0<p_T<100$ MeV/c pions, (b) and (d): $500<p_T<600$ MeV/c pions. In
(c) and (d) only pions with azimuthal direction in the range
$165^o<\phi<195^o$ (as indicated by the arrows) are shown (From
Ref.[26]).}
\end{figure}

\begin{figure}[th]
\centerline{\psfig{file=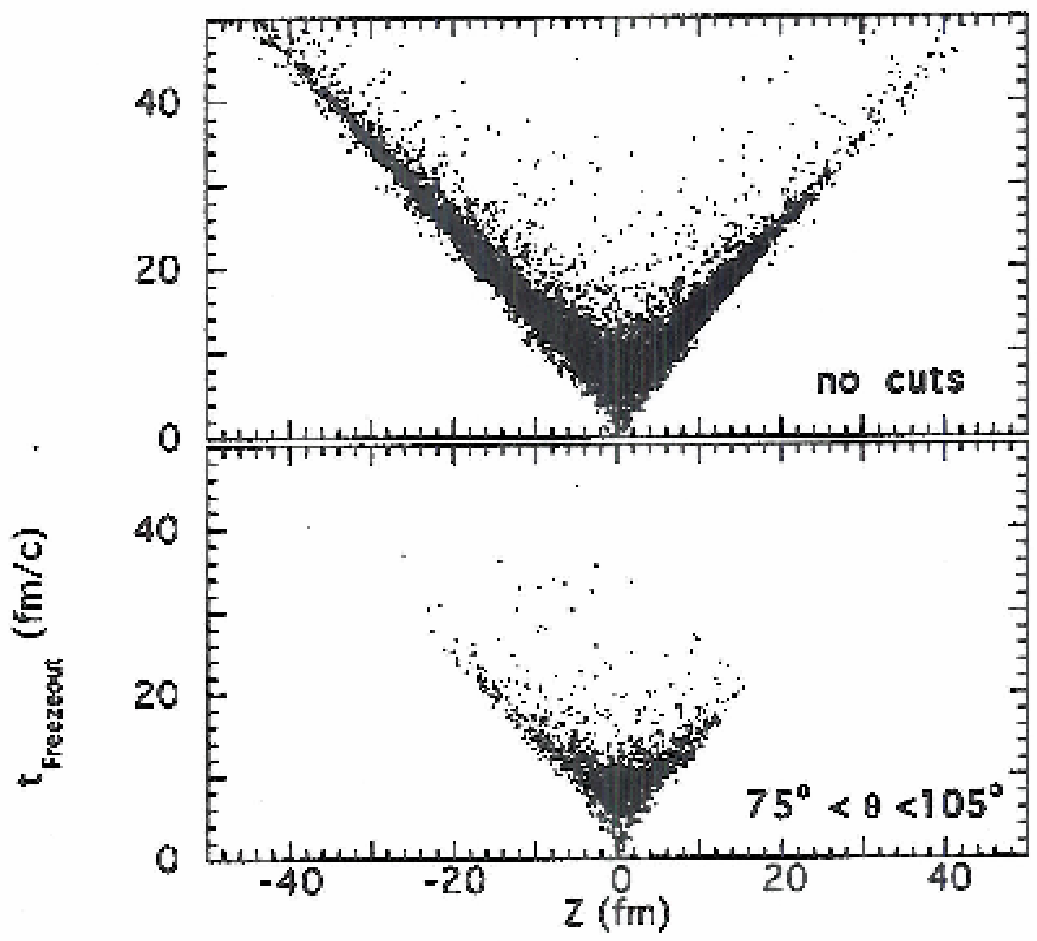,width=10cm}} \vspace*{8pt}
\caption{Distributions of pion freeze out positions for SPS $S+Pb$
collisions for time versus $z$. The upper plot has no cuts applied.
The lower plot only shows positions of pions with radial direction
in the range $75^o<\theta<105^o$ in the c.m. frame (corresponding to
a c.m. rapidity window of $\pm0.25$) (From Ref.[26]).}
\end{figure}

One can see the dramatic effect of rescattering on the freezeout
times of particles in the HRM calculation in Figure 14, which shows
the freezeout (and decay) time distributions for pions, kaons, and
nucleons for SPS Pb+Pb collisions (a) from a full rescattering
calculation, and (b) with rescattering turned off. With rescattering
the distributions for different particles are peaked at different
freezeout times, the kaons peaking earliest and the nucleons the
latest. Without rescattering the only feature to be seen is the
exponential decay of the initial resonances producing pions and
kaons.

\begin{figure}[th]
\centerline{\psfig{file=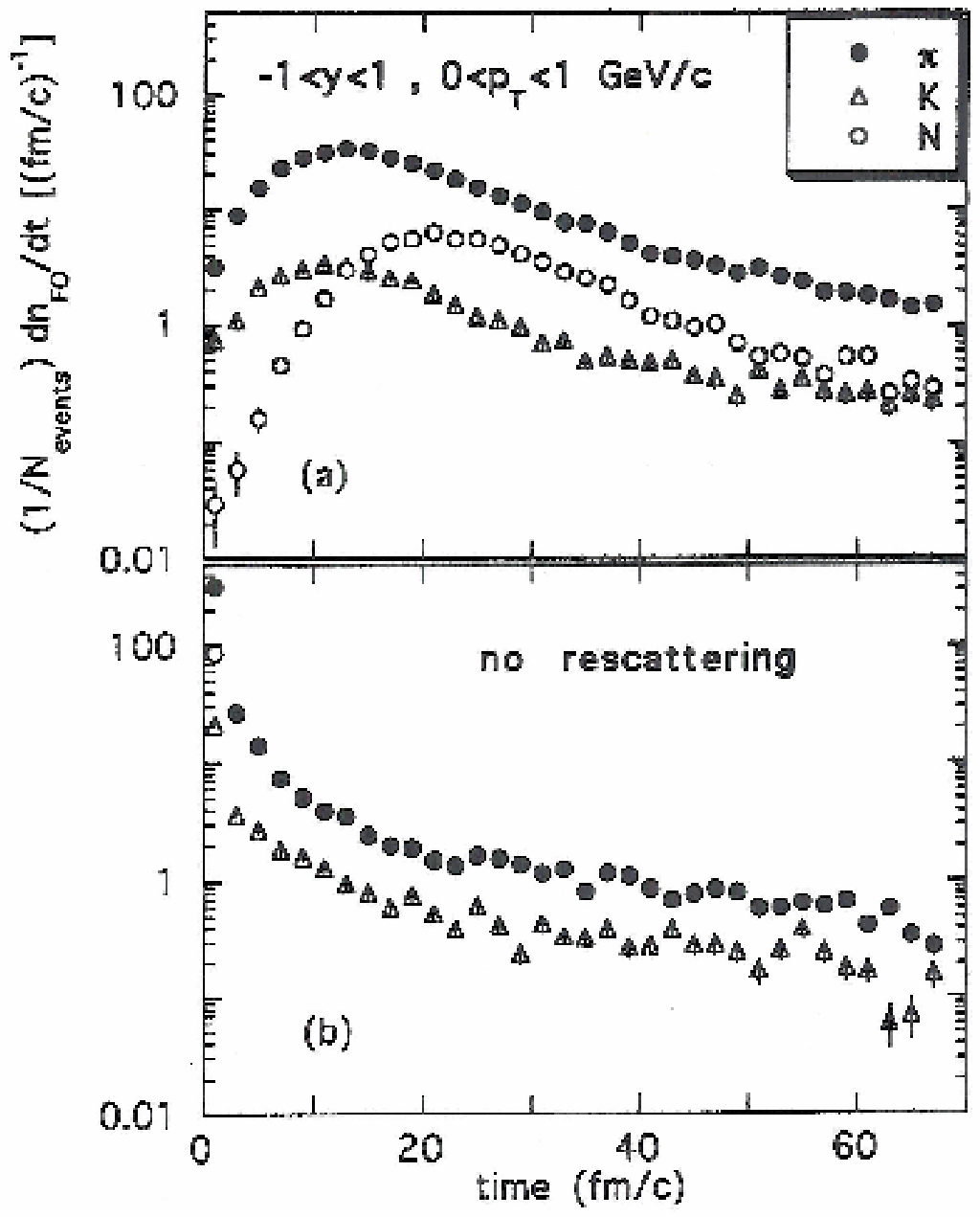,width=10cm}} \vspace*{8pt}
\caption{Freezeout (and decay) time distributions for pions, kaons,
and nucleons for SPS Pb+Pb (a) from a full rescattering calculation,
and (b) with rescattering turned off (From Ref.[21]).}
\end{figure}

Figures 15 and 16 show the effects of rescattering from the HRM
model on the transverse mass distributions for SPS Pb+Pb collisions
for pions, kaons, and nucleons. Figure 15 shows these distributions
at freeze out along with exponential fits from which slope
parameters are extracted using
\begin{equation}
\label{mt} \frac{1}{m_T}\frac{dN}{dm_T}=a\exp(-\frac{m_T}{T})
\end{equation}
where $m_T=\sqrt{{p_T}^2+m^2}$ is the transverse mass, $m$ is the
particle rest mass, $T$ is the slope parameter, and $a$ is a
normalization constant. Figure 16 shows how these slope parameters
evolve in time during the calculation. At $t=0$ fm/c all of the
particle types begin with a common temperature (as seen in Table
\ref{tab:tab1}) and as time proceeds the rescattering drives the
separation of the slope parameters until freezeout at which point
they evolve into the experimental values from the NA44
experiment\cite{na44} (as discussed above, $T_{init}$ is adjusted to
give the overall scale of the experimental slope parameters). Also
shown in Figure 16 is the time evolution for the pion slope
parameter for the ``pion gas'' case, i.e. without nucleons and kaons
included. As seen, the initial temperature required in this case is
much lower than for the full calculation, showing the significant
effect rescattering with the nucleons, and to a lesser extent kaons,
has on the pions.

\begin{figure}[th]
\centerline{\psfig{file=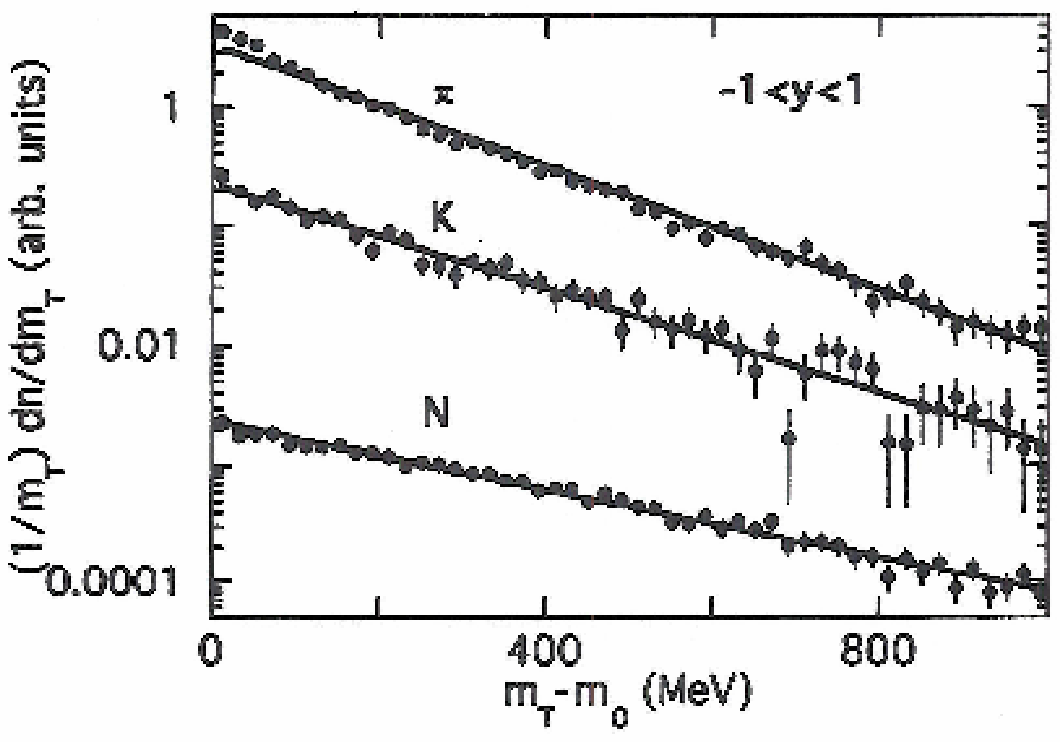,width=10cm}} \vspace*{8pt}
\caption{Transverse mass distribution at freeze out from HRM for SPS
Pb+Pb collisions for pions, kaons, and nucleons at midrapidity with
exponential fit (From Ref.[21]).}
\end{figure}

\begin{figure}[th]
\centerline{\psfig{file=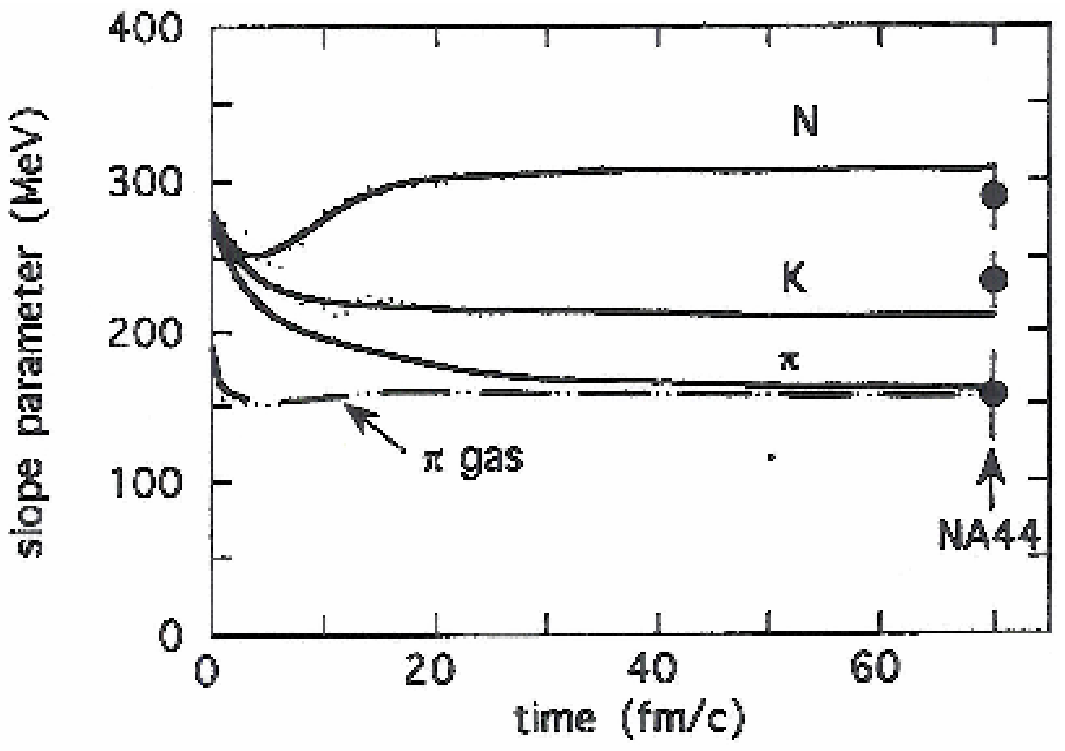,width=10cm}} \vspace*{8pt}
\caption{Time evolution of the slope parameters in HRM for SPS Pb+Pb
collisions for a full calculation (solid lines) and for a ``pion
gas'' calculation (dashed line). Experimental slope parameters from
NA44 (points) are plotted at $t=70$ fm/c for comparison (From Ref.
[21]).}
\end{figure}

Figure 17 shows projections of the three-dimensional two-pion
correlation function for SPS Pb+Pb onto the $Q_{Tside}$, $Q_{Tout}$,
and $Q_{Long}$ axes from HRM. A fit to Eq. (\ref{e9}) is also shown.
As seen, applying the boson symmetrization method described above
with the HRM results in correlation functions which are well
described by the gaussian model represented in Eq. (\ref{e9}). The
two extracted transverse radius parameters, $R_{Tside}$ and
$R_{Tout}$, are seen to be comparable in size around $5-6$ fm, while
$R_{Long}$ is significantly larger, reflecting the different
dynamics of the pion source along the beam direction compared with
transverse to the beam direction. Looking at the resulting $\lambda$
parameter extracted in this case of $0.634\pm0.019$, it is seen to
be significantly smaller than unity which was obtained in the INCM
discussed above. The explanation for this in the HRM model is the
presence of long-lived resonances such as $\eta$ and $\eta'$. Pions
produced from these resonances come from a much larger source that
the directly produced pions, resulting in an overall suppression of
the correlation function which is manifest in a smaller overall
$\lambda$ parameter for the collision. It is interesting to compare
this effect with the effect seen by Hanbury-Brown and Twiss when
observing Alpha Centauri, i.e, Eq. (\ref{alphac}). The two cases
turn out to be analogous since for Alpha Centauri one has two photon
sources, i.e. stars, separated by a large distance which dilutes the
HBT effect, whereas in RHI collisions the two pion sources separated
by a long distance are the direct and long-lived resonance sources.

\begin{figure}[th]
\centerline{\psfig{file=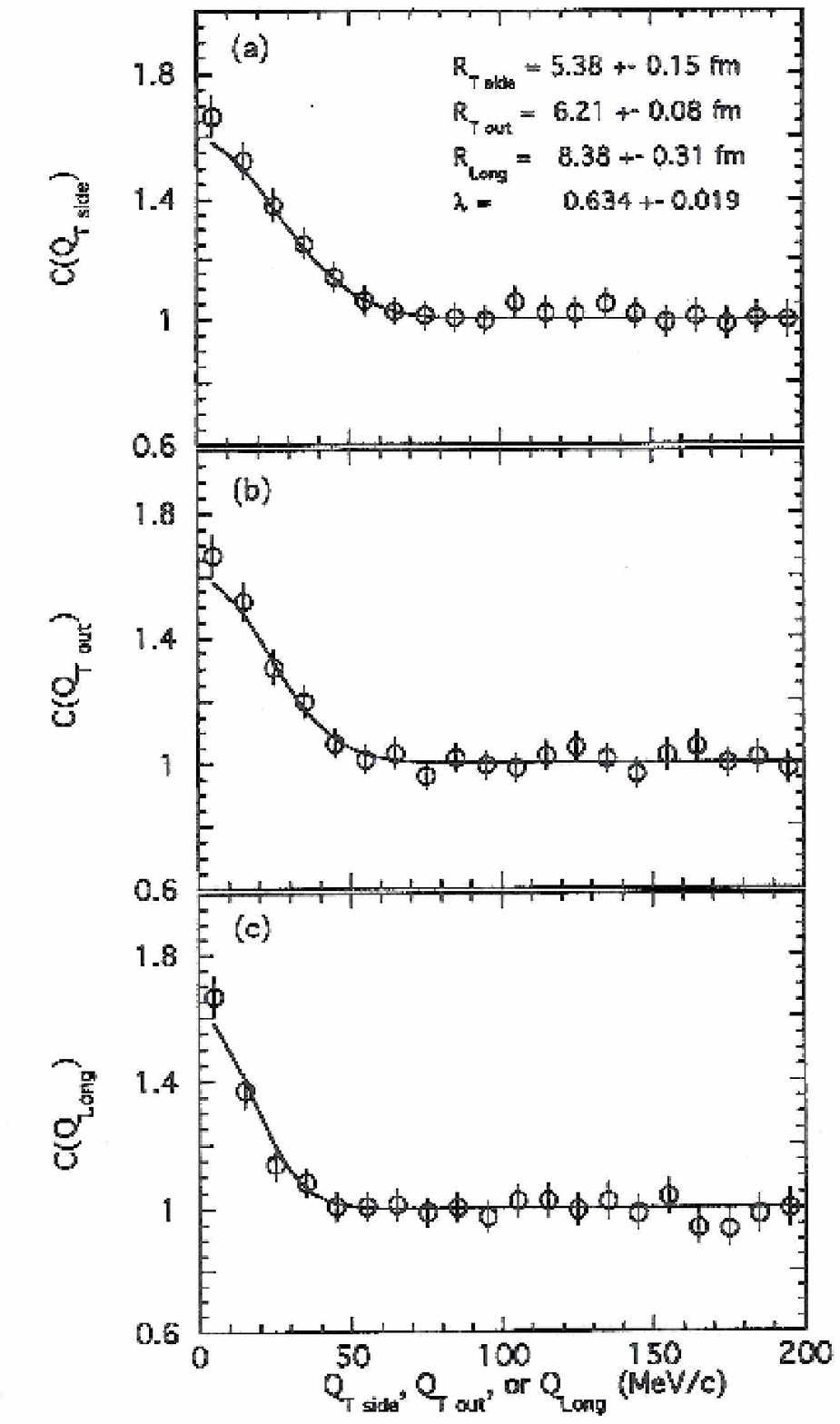,width=7cm}} \vspace*{8pt}
\caption{Projections of the three-dimensional two-pion correlation
function for SPS $Pb+Pb$ onto the $Q_{Tside}$, $Q_{Tout}$, and
$Q_{Long}$ axes from HRM. A fit to Eq. (\ref{e9}) is also shown
(From Ref.[21]).}
\end{figure}

It is interesting to compare at the time evolutions of various
observables in the context of the HRM for RHIC Au+Au collisions.
Figure 18 shows the time evolution of the pion elliptic flow, pion
HBT, and $m_T$ slope parameters from the rescattering calculation
with an impact parameter of 8 fm and averaged over 100 events. Lines
fitted to the points from the calculation are shown for convenience.
All quantities are extracted at midrapidity, elliptic flow includes
all $p_T$, HBT is calculated for $125<p_T<225$ MeV/c, and the slope
parameters are calculated in the region $m_T<1$ GeV. In these
calculations, rescattering drives all of the time evolution seen in
the various quantities, i.e. if rescattering were turned off, there
would be no change in their $t=0$ values for $t>0$. It is seen that
the elliptic flow develops the earliest, stabilizing at about 5
fm/c, the HBT parameters stabilizing next at about 10 fm/c, and the
latest being the slope parameters which require a time somewhat
longer than 25 fm/c to stabilize (note that rescattering
calculations are carried out to a time of 100 fm/c).

\begin{figure}[th]
\centerline{\psfig{file=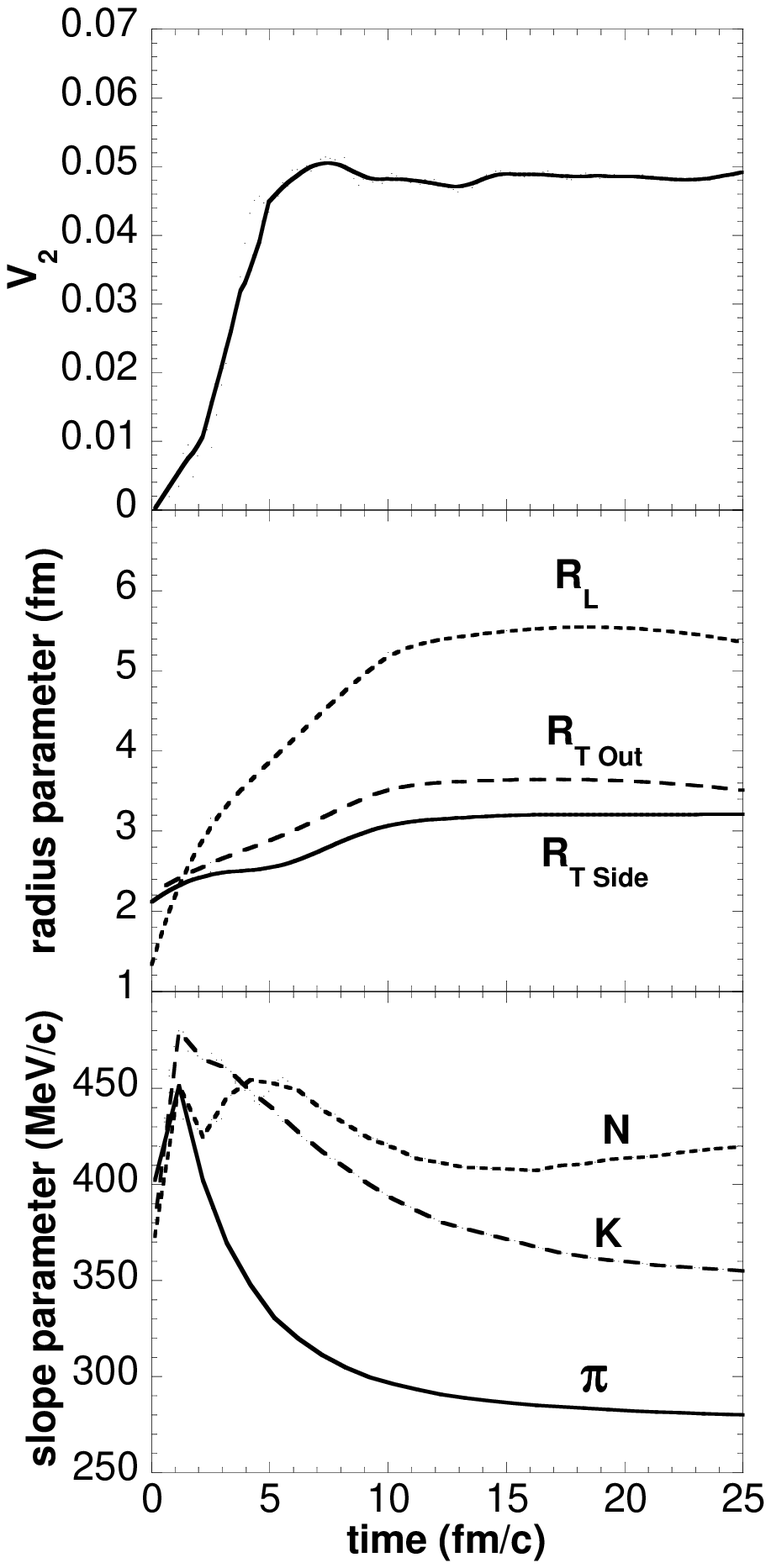,width=7cm}} \vspace*{8pt}
\caption{Time evolution of $V_2$, HBT, and $m_T$ slope parameters
from HRM for RHIC Au+Au.}
\end{figure}

\subsubsection{Subdivision test of HRM code}
Before presenting comparisons of the HRM code with experimental
results, it is worthwhile to address two criticisms which have been
made against using the HRM to make RHIC-energy predictions. They are
1) the initial state for the calculation is too hot and too dense to
be considering hadrons, thus the results are meaningless, and 2) the
calculational results may have reasonable agreement with data but
that is only accidental because the calculation is dominated by
computational artifacts which strongly influence the results. A
response to 1) is that the HRM should be viewed as a limiting case
study of how far one can get with an extreme and simple model such
that maybe we can learn something about the true initial state from
this unexpected agreement with data, e.g. maybe hadron-like objects
exist in the QGP, and/or the QGP has a short lifetime and then
quickly hadronizes. The response to 2) is that the HRM has been
tested for Boltzmann-transport-equation-like behavior and the
influence of superluminal artifacts using the subdivision
method.\cite{molnar2000,humanic2003a} Although in that test the HRM
results were shown to not be significantly affected by using a
subdivision of $l=5$, it was not studied whether $l=5$ was
sufficiently large to significantly reduce the superluminal
artifacts to make the test meaningful. Superluminal artifacts can be
introduced into a scattering code at the point at which the
scattering cross section is used to determine whether two particles
are sufficiently close in space and time for a scattering to take
place if the density of particles in the calculation is sufficiently
high.\cite{molnar2000} In a subdivision test of a scattering code,
the particle density, $\rho$, is increased by a factor $l$, the
subdivision, while at the same time the scattering cross section,
$\sigma$, is decreased by this same factor, i.e.
\begin{eqnarray}
\label{subdiv} \rho \rightarrow l\rho\\
\sigma \rightarrow \frac{\sigma}{l}.\nonumber
\end{eqnarray}
Running the scattering code in this configuration with $l>1$ in
principle reduces the superluminal artifacts and tests whether the
code is properly solving the Boltzmann transport equation.

In the present test, subdivisions of $l=1, 5,$ and $8$ are used in
HRM calculations of RHIC-energy Au+Au collisions with $b=8$ fm
centrality. Plots of the transverse signal velocity distributions
for these subdivision are shown in order to determine how effective
these subdivisions are in reducing the superluminal effects. The
pion HBT radius parameters are also shown for these subdivisions in
order to determine how sensitively they depend on them. The left
plot in Figure 19 shows the transverse signal velocity distributions
for all particles in the calculation. As shown, there are indeed
superluminal effects present for $l=1$, but the higher subdivisions
significantly reduce these effects. From this, it is seen that using
$l=5$ and $l=8$ should each provide a valid test of the effects of
these artifacts on the results of the calculation. Results from
calculating pion HBT parameters vs. $p_T$ for $b=8$ fm is shown in
the right plot of Figure 19. As is seen, the higher subdivisions do
not significantly effect the HBT results. Radial and elliptic flow
results from the HRM calculations can also be shown to not be
effected significantly by using these higher
subdivisions.\cite{humanic2003a} Thus one can conclude from this
test that the previously published results and present results from
the HRM are not affected by superluminal artifacts, and criticism 2)
above is answered. One can now proceed with comparisons with data
below.

\begin{figure}[b]
\hspace{.1in} \scalebox{.9}{\includegraphics{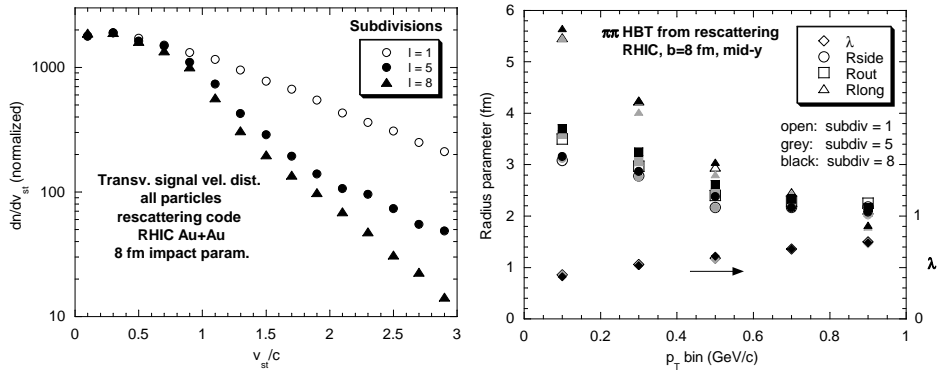}}
\caption{Subdivision test of the HRM code, showing transverse signal
velocity distribution and pion HBT parameters vs. $p_T$ for $b=8$ fm
centrality and subdivisions 1, 5, and 8.} \label{fig:hbt1}
\end{figure}

\subsection{Comparison of HRM with AGS ($\sqrt{s} \sim 2$ GeV/nucleon) and SPS
 ($\sqrt{s} \sim 20$ GeV/nucleon) experiments}
Having discussed some of the features of the HRM, comparisons
between HBT results from the HRM with AGS and SPS experiments are
now presented below in Figures 20-23.

Figure 20 shows a comparison between boson source parameters from
the HRM (triangles) with those from the fixed-target central
collisions 200 GeV/nucleon S+Pb measured by SPS experiment
NA44\cite{bek,na44b} and 14.6 GeV/nucleon Si+Au from the AGS
experiment E859/E866\cite{ags} (circles). Both experiments are based
on small acceptance magnetic spectrometers with good particle
identification allowing them both to also carry out HBT measurements
with kaon pairs as well as pion pairs as shown in the figure. As
seen, NA44 also measured pion source parameters with a low-$p_T$ cut
($<p_T>\sim 150$ MeV/c) and a high-$p_T$ cut ($<p_T>\sim 450$ MeV/c)
on the pion momentum. The HRM is seen to follow the trends of the
data rather well, predicting the decrease in the source parameters
with the higher pion momentum cut (NA44) and for both experiments
predicting smaller source parameters extracted from the kaon-pair
HBT measurements. In the context of the HRM, the explanation for the
smaller parameter sizes extracted used kaon pairs is due to the
generally smaller scattering cross sections for $K-\pi$ and $K-N$
reactions compared with those for pions. The dashed lines in the
figure show the parameters extracted from HRM with rescattering
turned off, showing that rescattering plays a crucial role in
determining the scale and kinematic dependencies of the boson source
parameters measured by HBT interferometry.

\begin{figure}[th]
\centerline{\psfig{file=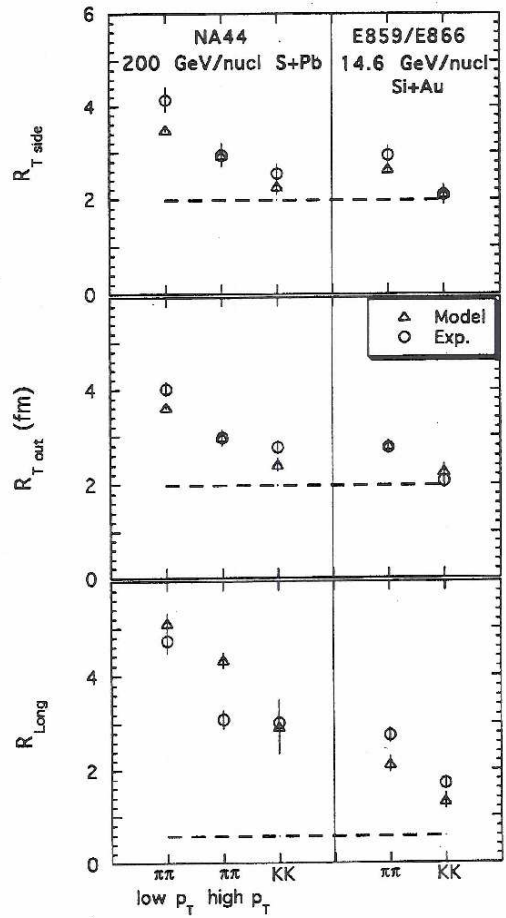,width=7cm}} \vspace*{8pt}
\caption{Comparison between boson source parameters from the HRM
(triangles) with those from SPS experiment NA44 and AGS experiment
E859/E866 (circles). The dashed lines show the parameters extracted
from HRM with rescattering turned off (From Ref.[26]).}
\end{figure}

Figure 21 presents a comparison between the HRM for various model
configurations and HBT parameters extracted from fixed-target
central 158 GeV/nucleon Pb+Pb collisions from SPS experiment NA49
(based on a large acceptance detector).\cite{na49} The dashed lines
are projections of the NA49 data points to guide the eye. For the
purposes of this figure, ``IOC'' refers to the use of Eq.
(\ref{rel}) in the HRM (the normal method used) and ``pill'' refers
to taking $t_{had}=0$ and $z_{had}$ uniformly distributed in the
region $z=\pm1$. As seen, the HRM predictions show robustness to the
first three running configurations predictions being close to the
measurements for the cases ``IOC'', ``pill'', and ``pill(old)'', the
latter being a calculation without resonances which accounts for the
$\lambda$ parameter being unity for that case. The cases ``IOC (pion
gas)'' and ``IOC (no RS)'', referring to running the HRM code
without kaons and nucleons and with rescattering turned off,
respectively, are seen to deviate significantly from the
measurements showing the importance of including the pions and kaons
and rescattering in the calculations.

\begin{figure}[th]
\centerline{\psfig{file=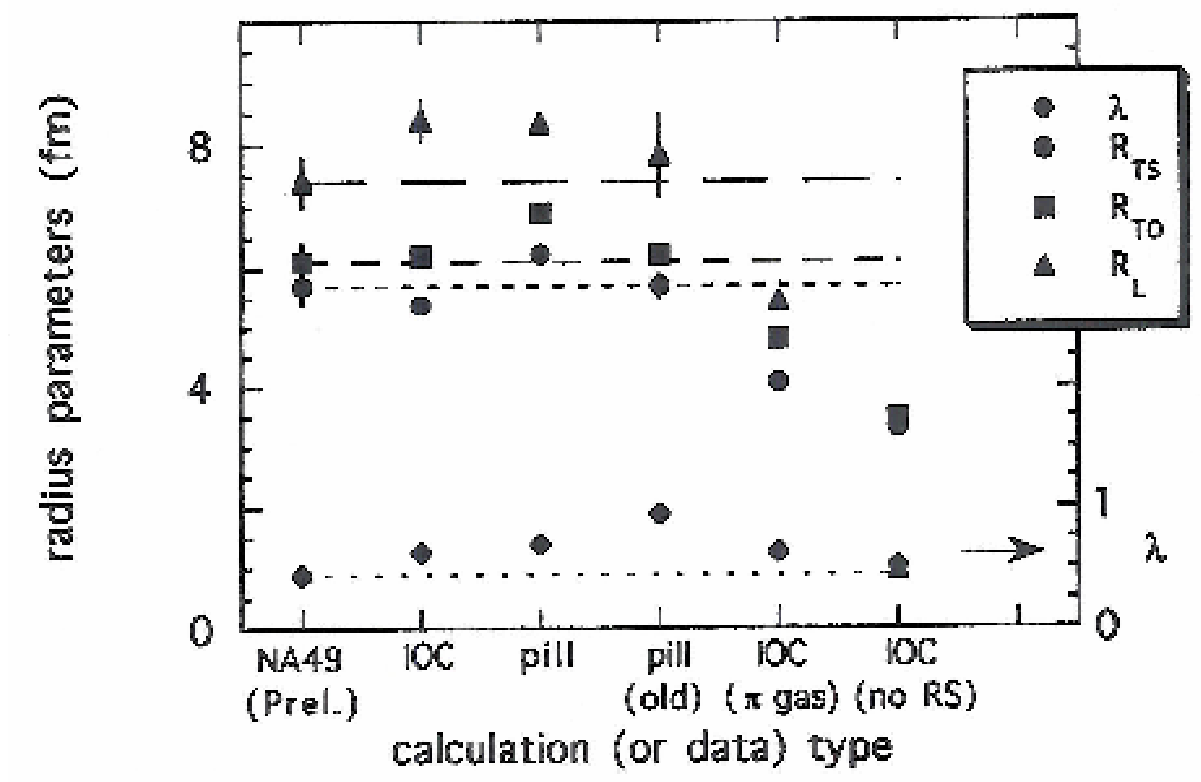,width=10cm}} \vspace*{8pt}
\caption{Comparison between SPS experiment NA49 HBT results for
$Pb+Pb$ with HRM for various model configurations. The dashed lines
are projections of the NA49 data points to guide the eye (From
Ref.[21]).}
\end{figure}

Figures 22 and 23 show more comparisons of HBT source parameters
extracted from the HRM and the SPS NA49 and NA44 experiments. Figure
22 shows a comparison between HRM (points) and the trends of NA49
central 158 GeV/nucleon Pb+Pb experimental results\cite{na49}
(dashed lines with same meaning as in Figure 21) for the $k_T$
dependence of the pion source parameters. Figure 23 compares an
overview of HBT-extracted pion and kaon source parameters measured
in the NA44 experiment\cite{bek,fran,bear2,humanic1999} with those
calculated with the HRM. Although there are some minor
disagreements, the HRM is seen overall to qualitatively describe the
trends of the data for both experiments.

\begin{figure}[th]
\centerline{\psfig{file=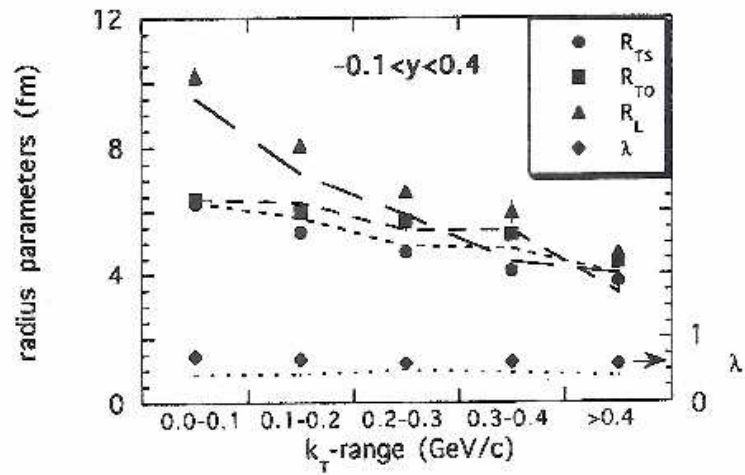,width=10cm}} \vspace*{8pt}
\caption{Comparison between HRM (points) and the trends of NA49
$Pb+Pb$ experimental results (dashed lines with same meaning as in
Figure 21) for the $k_T$ dependence of the pion source parameters
(From Ref.[21]).}
\end{figure}

\begin{figure}[th]
\centerline{\psfig{file=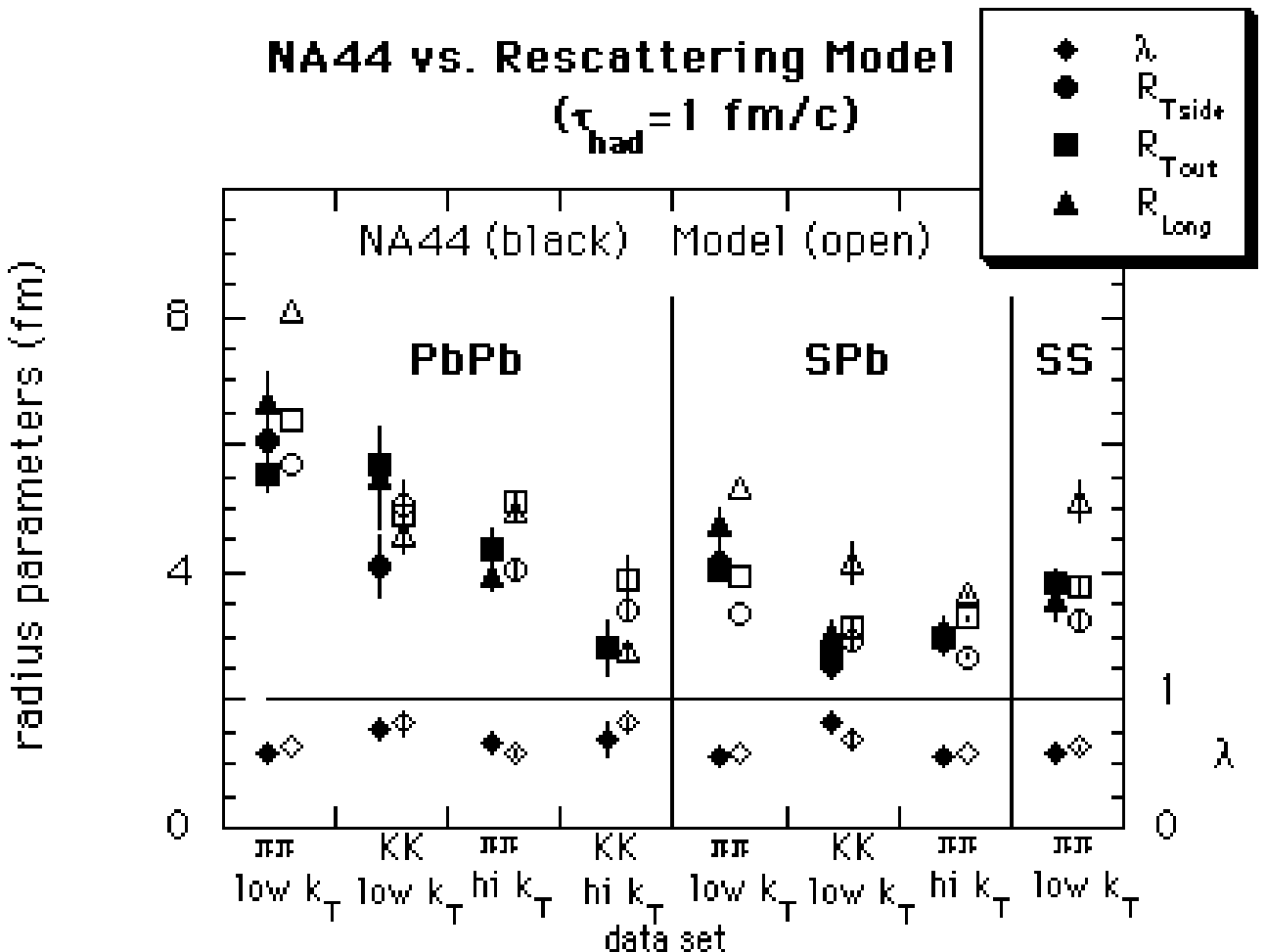,width=10cm}} \vspace*{8pt}
\caption{Comparison between HRM calculations and boson source
parameters extracted from the NA44 experiment (From Ref.[36]).}
\end{figure}

\subsection{Comparison of HRM with RHIC ($\sqrt{s} \sim 130-200$ GeV/nucleon) experiments}
Results from the first year of running of the Relativistic Heavy Ion
Collider (RHIC) for Au+Au collisions at $\sqrt{s}=130$ GeV showed
surprisingly large pion elliptic flow\cite{Adler:2001a} and
surprisingly small radii from two-pion HBT interferometry.
\cite{Adler:2001b,Adcox:2002a} Hydrodynamical models agreed with the
large elliptic flow seen in the RHIC data\cite{Kolb:1999a} but
significantly disagreed with the experimental HBT radii.
\cite{Rischke:1996a} On the other hand, relativistic quantum
molecular dynamics calculations which include hadronic rescattering,
for example RQMD v2.4,\cite{Sorge:1989a} significantly under-predict
the elliptic flow seen in the RHIC data\cite{Sorge:1995a} but
predict pion HBT radii comparable to the data.\cite{Hardtke:1999a} A
calculation was made to extract HBT radii with a hydrodynamical
model coupled with a hadronic rescattering afterburner with the
result that the HBT radii were significantly larger than
measurements.\cite{Soff:2000a} This situation lead to the first big
mystery from RHIC, sometimes called the ``HBT Puzzle.'' It has even
been suggested that one should call into question the current
understanding of what information pion HBT measurements
give.\cite{Gyulassy:2001a} In this context, the HRM model is
compared with both elliptic flow and HBT measurements from RHIC to
see if it can shed any light on this situation.

\subsubsection{Elliptic flow results}
 The elliptic flow variable for a collision, $V_2$, is
defined as\cite{Adler:2001a}
\begin{eqnarray}
\label{v2} V_2=<\cos(2\phi)> \\\nonumber
    \phi=\arctan(\frac{p_y}{p_x})
\end{eqnarray}
where $p_x$ and $p_y$ are the $x$ and $y$ components of the particle
momentum, and $x$ is in the impact parameter direction, i.e.
reaction plane direction, and $y$ is in the direction perpendicular
to the reaction plane.

Figure 24 shows the $p_T$ dependence of $V_2$ for pions and nucleons
extracted from the $b=8$ fm HRM calculation compared with the trends
of the STAR measurements for $\pi^{+}+\pi^{-}$ and $p$ +
$\overline{p}$ at $11-45\%$ centrality,\cite{Adler:2001a} which
roughly corresponds to this impact parameter.

\begin{figure}
\centerline{\psfig{file=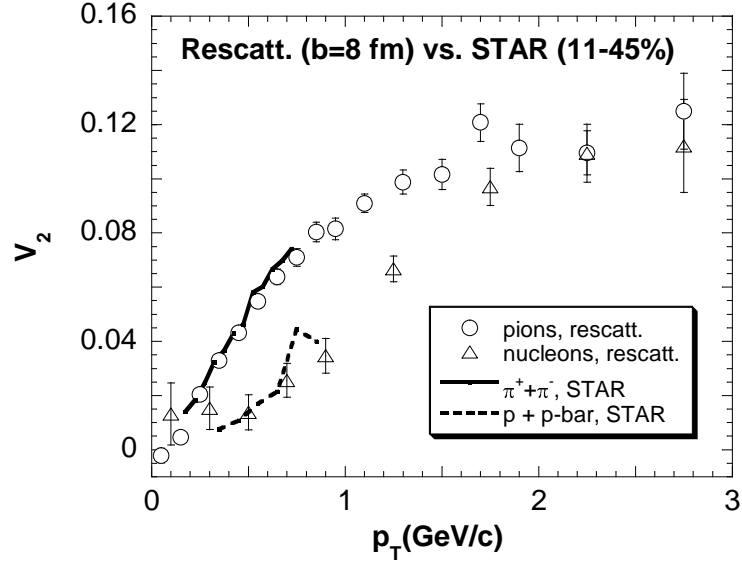,width=10cm}} \vspace*{8pt}
\caption{\label{fig:v2} Calculations of $V_2$ from the HRM for $b=8$
fm for pions and nucleons compared with STAR measurements at 11-45\%
centrality. The plotted points with error bars are the HRM
calculations and the lines show the trends of the STAR measurements.
Average errors on the STAR measurements are $\leq0.002$ for pions
and $0.006$ for protons+antiprotons (From Ref.[22]).}
\end{figure}

Figure 25 compares the $p_T$ dependence of $V_2$ for kaons from the
$b=8$ fm HRM calculation with the STAR measurements for ${K^0}_s$ at
$11-45\%$ centrality.\cite{Adler:2002b} As seen, the HRM calculation
values are in reasonable agreement with the STAR measurements. The
flattening out of the pion and nucleon $V_2$ distributions for
$p_T>2$ GeV/c is consistent with that seen in STAR and PHENIX
results for minimum-bias hadrons\cite{Snellings:2001a,Zajc:2001a}
(the kaon $V_2$ calculation does not extend higher than 2 GeV/c in
Figure 5 due to limited statistics). Thus, the same rescattering
mechanism that can account for the radial flow seen in $m_T$
distributions, e.g. Figures 15 and 16, also is seen to account for
the magnitude and $p_T$ dependence of the elliptic flow for pions,
kaons, and nucleons.

\begin{figure}
\centerline{\psfig{file=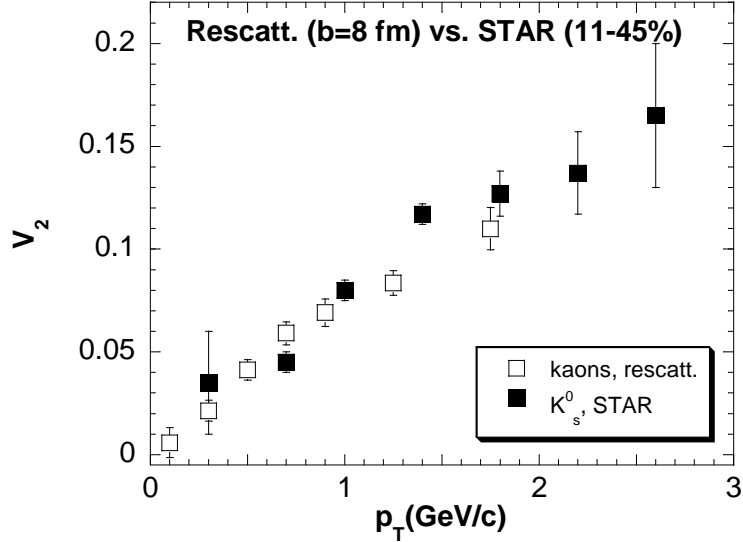,width=10cm}} \vspace*{8pt}
\caption{\label{fig:kv2} Calculation of $v_2$ from HRM for kaons at
$b=8$ fm compared with STAR measurements for ${K^0}_s$ at 11-45\%
centrality (From Ref.[22]).}
\end{figure}

\subsubsection{``Year-1'' HBT results}
The pion source parameters extracted from HBT analyses of HRM
calculations for three different impact parameters, $b=0$, $5$, and
$8$ fm, are compared with STAR $\pi^{-}$ measurements at three
centrality bins\cite{Adler:2001b} in Figure 26. Note that the PHENIX
experiment HBT results\cite{Adcox:2002a} are in basic agreement with
the STAR results. The STAR centrality bins labeled ``3'', ``2'', and
``1'' in the figure correspond to $12\%$ of central, the next
$20\%$, and the next $40\%$, respectively. These bins are roughly
approximated by the impact parameters used in the HRM calculations,
i.e. the average impact parameters of the STAR centrality bins are
estimated to be within $\pm2$ fm of the HRM calculation impact
parameters used to compare with them. In the left panel, the
centrality dependence of the HBT parameters is plotted for a $p_T$
bin of $0.125-0.225$ GeV/c. In the right panel, the $m_T$ dependence
of the HBT parameters is plotted for centrality bin 3, for the STAR
measurements, or $b=0$ fm, for the HRM calculations. Although there
are differences in some of the details, the trends of the STAR HBT
measurements are seen to be described rather well by the HRM
calculation.

\begin{figure}
\centerline{\psfig{file=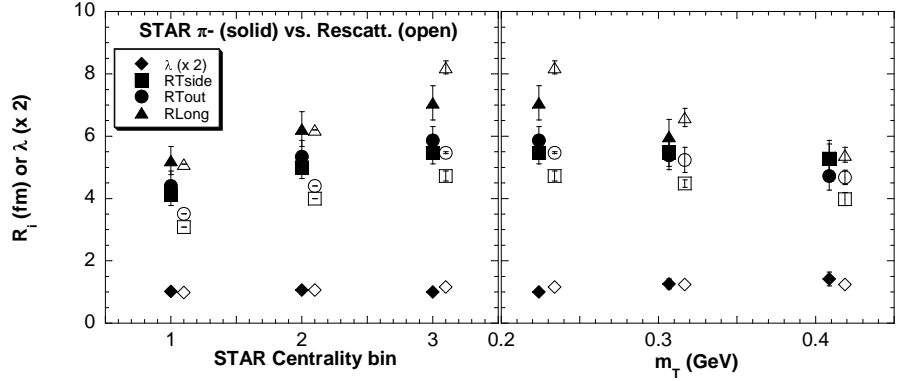,width=12cm}} \vspace*{8pt}
\caption{\label{fig:hbt} Comparison of HBT source parameters from
HRM with STAR measurements as a function of centrality bin (see
text) and $m_{T}$. The STAR measurements are the solid symbols and
the rescattering calculations are the open symbols. The errors on
the STAR measurements are statistical+systematic (From Ref.[22]).}
\end{figure}

\subsubsection{Azimuthal HBT results}
It was first shown experimentally at the AGS that looking at the
azimuthal dependence of HBT with respect to the reaction plane can
give another handle on the space-time evolution of the pion
source.\cite{e895} For central collisions with vanishing impact
parameter ($b\sim0$) one would expect no azimuthal dependence since
the initial collision is symmetric about the beam axis. However, for
non-central ($b>0$) collisions, a definite initial asymmetry with
respect to the beam axis will exist which may be reflected in the
extracted HBT source parameters. The STAR collaboration has recently
published experimental results on the azimuthal dependence of pion
HBT parameters in central and non-central RHIC
collisions.\cite{star2004} Preliminary calculations have been made
with the HRM to extract azimuthal HBT parameters to compare with the
STAR results. These comparisons are shown in Figures 27 and 28
below. In Figure 27 the azimuthal dependence of HBT parameters is
shown for $b=0$ fm rescattering model calculations compared with
results from STAR central collisions ($0-5\%$ centrality). As seen
for both the calculation and data, no oscillations occur with
respect to $\phi$ for any of the parameters except the cross-term
parameter, $R_{outside}^2$, as would be expected for an azimuthally
symmetric system. Figure 28 shows a similar comparison for $b=4$ fm
calculations and STAR medium-central collisions ($10-20\%$
centrality). As would be expected, the non-central collisions break
the azimuthal symmetry of the pion source and oscillations are now
seen in all parameters. The calculations are seen to be in
reasonable qualitative agreement with the data.

\begin{figure}
\hspace{.1in} \scalebox{1.}{\includegraphics{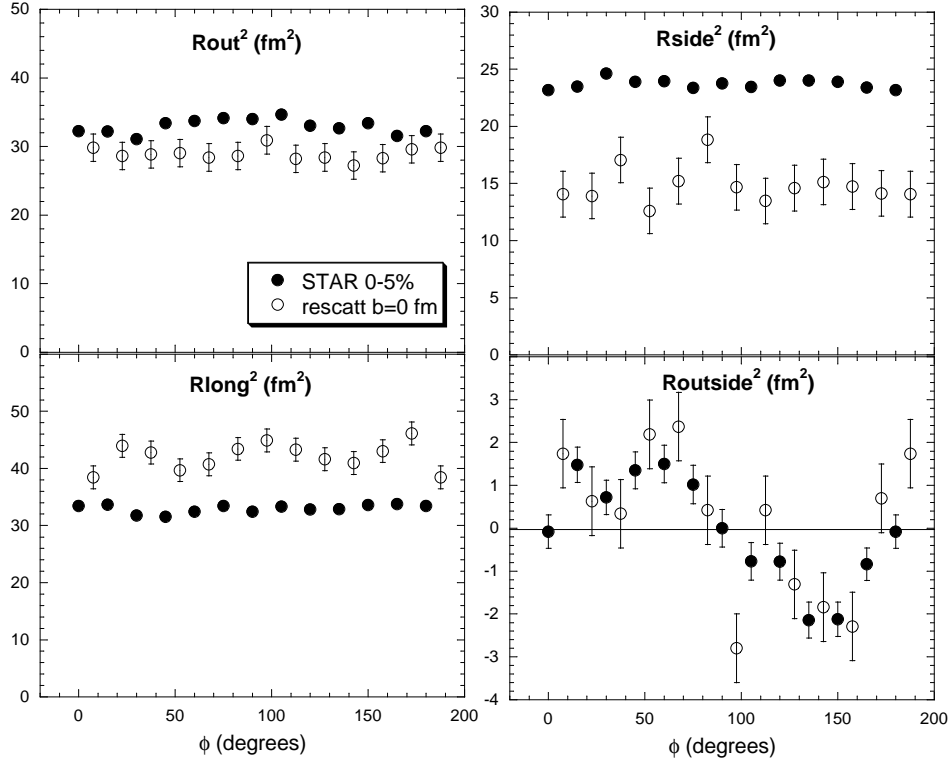}}
\caption{Azimuthal pion HBT parameters from the rescattering model
with $b=0$ fm centrality compared with central STAR results }
\label{fig:hbt2}
\end{figure}

\begin{figure}
\hspace{.1in} \scalebox{1.}{\includegraphics{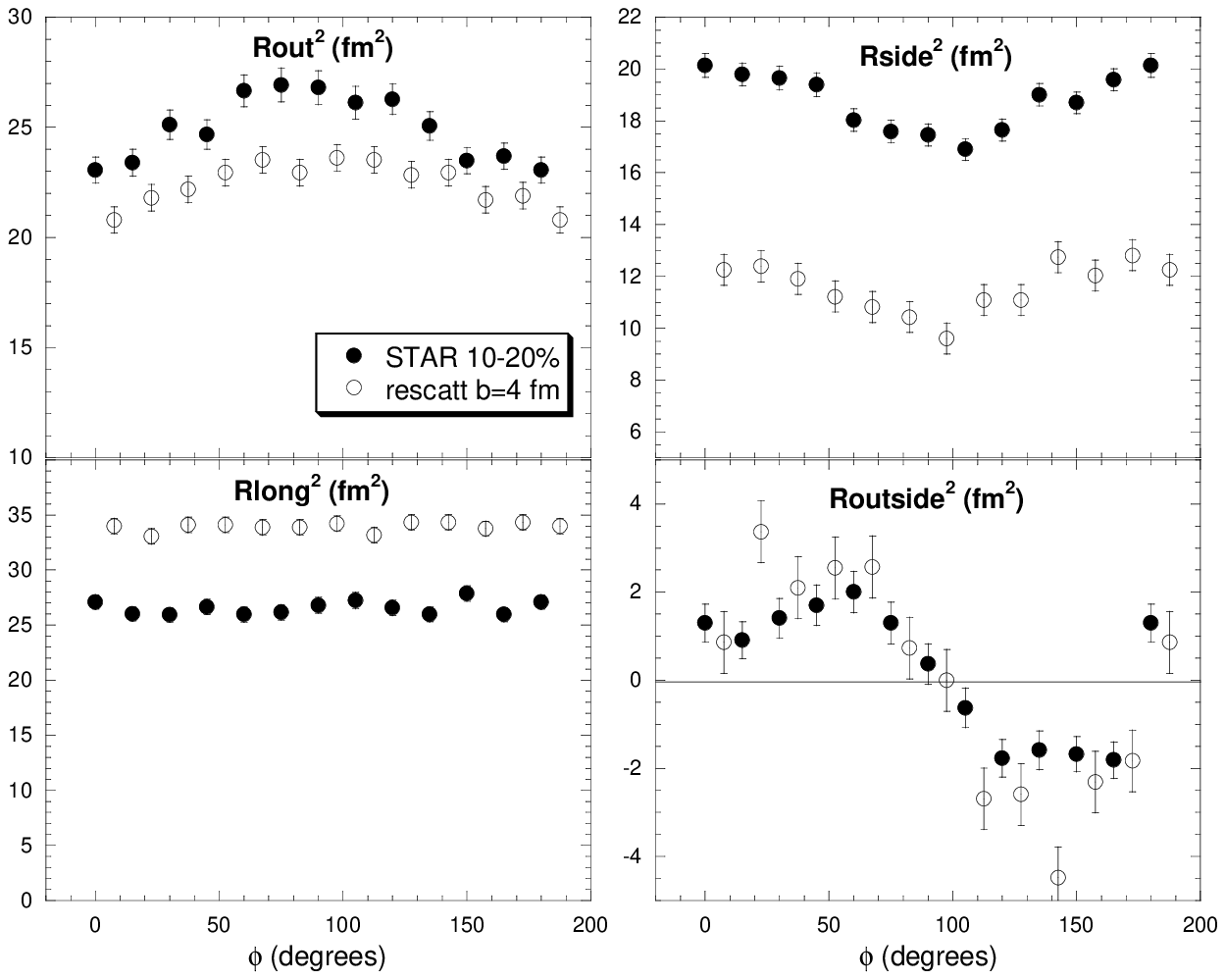}}
\caption{Azimuthal pion HBT parameters from the rescattering model
with $b=4$ fm centrality compared with medium-central STAR results }
\label{fig:hbt3}
\end{figure}

\subsubsection{Discussion of RHIC results}
As shown above, the elliptic flow as well as the features of the HBT
measurements at RHIC can be adequately described by HRM with the
hadronization model parameters given earlier in Table I. The results
of the calculations are found to be sensitive to the value of
$\tau_{had}$ used, as was studied in detail for SPS HRM
calculations.\cite{humanic1998} For calculations with $\tau_{had}>1$
fm/c the initial hadron density is smaller, fewer collisions occur,
and the rescattering-generated flow is reduced, reducing in
magnitude the radial and elliptic flow and most of the HBT
observables. Only the HBT parameter $R_{Long}$ increases for larger
$\tau_{had}$ reflecting the increased longitudinal size of the
initial hadron source, as seen in Eq. (\ref{rel}). One can
compensate for this reduced flow in the other observables by
introducing an ad hoc initial ``flow velocity parameter'', but the
increased $R_{Long}$ cannot be compensated by this new parameter. In
this sense, the initial hadron model used in the present
calculations with $\tau_{had}=1$ fm/c and no initial flow is
uniquely determined with the help of $R_{Long}$.

At this point, one can consider the physical significance of the HRM
in two different ways. The first way is to accept that it is
physically valid in the time range where hadronic rescattering
should be valid, e.g. for times later than when the particle density
reaches about 1 $fm^{-3}$, and to take the initial state
hadronization model as merely a parameterization useful to fit the
data. Considering the calculation this way, one can at least expect
to gain an insight into the phase-space configuration of the system
relatively early in the collision (in the calculation 1 $fm^{-3}$
occurs at a time 4 fm/c after hadronization). The second way to
consider the calculation is to see if it is possible to also
physically motivate the initial state hadronization model. An
attempt to do this is given below.

In order to consider the present initial state model as a physical
picture, one must assume: 1) hadronization occurs very rapidly after
the nuclei have passed through each other, i.e. $\tau_{had}=1$ fm/c,
2) hadrons or at least hadron-like objects can exist in the early
stage of the collision where the maximum value of $\rho$ approaches
8 GeV/$fm^3$, and 3) the initial kinetic energies of hadrons can be
large enough to be described by $T=300$ MeV in Equation 1.

Addressing assumption 2) first, in the calculation the maximum
number density of hadrons at mid-rapidity at $t=0$ fm/c is 6.8
$fm^{-3}$, rapidly dropping to about 1 $fm^{-3}$ at $t=4$ fm/c.
Since most of these hadrons are pions, it is useful as a comparison
to estimate the effective volume of a pion in the context of the
$\pi-\pi$ scattering cross section, which is about 0.8 $fm^2$ for
s-waves.\cite{Prakash:1993a} The ``radius'' of a pion is found to be
0.25 fm and the effective pion volume is 0.065 $fm^3$, the
reciprocal of which is about 15 $fm^{-3}$. From this it is seen that
at the maximum hadron number density in the calculation, the
particle occupancy of space is estimated to be less than $50\%$,
falling rapidly with time. One could speculate that this may be
enough spacial separation to allow individual hadrons or hadron-like
objects to keep their identities and not melt into quark matter,
resulting in a ``super-heated'' semi-classical gas of hadrons at
very early times, as assumed in the present calculation.

Since the calculation takes the point of view of being purely
hadronic, it is instructive to consider assumptions 1) and 3) in the
context of the Hagedorn thermodynamic model of hadronic collisions.
\cite{Hagedorn:1970a} According to Hagedorn, the mass spectrum of
hadrons of mass m increases proportional to $\exp{(m/T_0)}$ in
hadronic collisions, where $T_0=160$ MeV is the limiting temperature
of the system. This seems to contradict the value $T=300$ MeV needed
in the present case in Eq. (\ref{temp}) to describe the data. The
Hagedorn model assumes that a) the system comes to equilibrium and
b) the details of particle production via direct processes and
through resonance decay average out. Neither of these assumptions is
necessarily guaranteed at very early times in the collision. The use
of the thermal functional form, Equation 1, to set up the initial
transverse momenta of the hadrons in the present calculations is
convenient but not required. For example, the exponential form
$\exp{(-m_T/T_e)}$ (where $T_e$ is a slope parameter) which does not
describe thermal equilibrium, could have equally well been used.
This exponential form of the transverse mass distribution was
successfully used previously in rescattering calculations to
describe SPS data.\cite{humanic1996,humanic1994}

Assumption 1) can also be motivated by the Color Glass Condensate
model.\cite{McLerran:2002a,Kovner:1995a} In the usual version of
this picture, after the collision takes place the Color Glass melts
into quarks and gluons in a timescale of about 0.3 fm/c at RHIC
energy, and then the matter expands and thermalizes into quark
matter by about 1 fm/c. In the context of the HRM calculations, it
is tempting to modify the collision scenario such that instead of
the Color Glass melting into quarks and gluons just after the
collision, the sudden impact of the collision ``shatters'' it
directly into hadronic fragments on the same timescale as in the
parton scenario due to the hadronic strong interactions.

\subsection{Predictions from HRM for LHC ($\sqrt{s} \sim 5500$ GeV/nucleon) experiments}
Since it has been found that the predictions from HRM agree rather
well with AGS, SPS, and RHIC measurements, it is interesting to use
this model to make similar predictions for Pb+Pb collisions at the
LHC. Preliminary calculations for the LHC have been carried out with
HRM, the results from which are shown below. In performing LHC
calculations, the following parameters were used in the code: 1) a
collision impact parameter of $b=8$ fm, 2) an initial temperature
parameter of $500$ MeV, 3) a hadronization proper time for the
initial system of 1 fm/c, 4) a $dn/dy$ at mid-rapidity for central
collisions for all particles of 4000, and 5) an initial rapidity
width of 4.2. These parameters were judged to be reasonable guesses
to simulate LHC Pb+Pb collisions. They at least satisfy the
self-consistency check that summing over the energy of all particles
in an event at the end of the calculation agrees with the input
total energy of a LHC Pb+Pb collision with an impact parameter of
$b=8$ fm. An impact parameter of $b=8$ fm was chosen for the present
preliminary study both to obtain non-negligible elliptic flow values
and for calculational convenience (even for this impact parameter
the cpu time used by the code for each LHC Pb+Pb event was about 60
hours). For item 3) above, the hadronization proper time was taken
to be the same as was used in the SPS and RHIC calculations. Results
of these calculations are compared with similar calculations at
$b=8$ fm centrality for RHIC Au+Au collisions and are shown in
Figures 29, 30, and 31. All of these results are obtained at
mid-rapidity, i.e $-2<y<2$.

\begin{figure}[b]
\hspace{.4in}
\scalebox{0.9}{\includegraphics{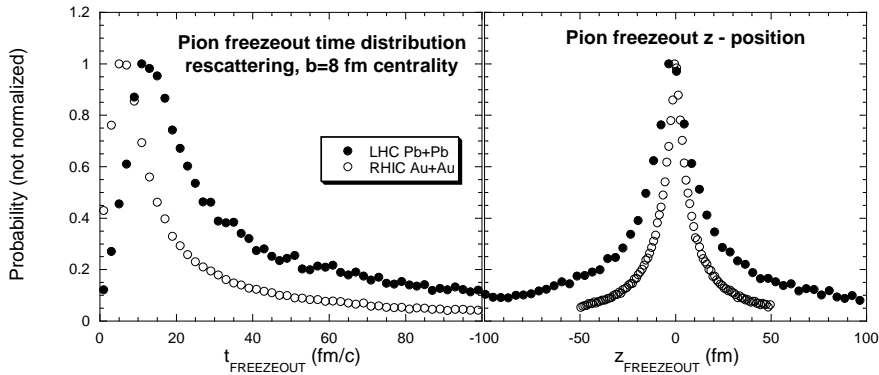}}\caption{Pion freezout
time and z-position distributions for LHC Pb+Pb and RHIC Au+Au for
$b=8$ fm centrality collisions at mid-rapidity from HRM.}
\label{fig:hbt1}
\end{figure}

Figure 29 shows the pion freezeout time and z-position distributions
for LHC Pb+Pb and RHIC Au+Au from HRM. As seen, the average pion
freezeout time and z-position for LHC are about twice as large as
those for RHIC. Although the tails of the freezeout time
distributions extend beyond 100 fm/c, the peaks for the LHC and RHIC
occur at fairly short times in the collision, at about 5 fm/c and 10
fm/c, respectively. Thus, effects from earlier times in general have
the greatest influence on the results from these calculations.

\begin{figure}
\hspace{.4in} \scalebox{0.9}{\includegraphics{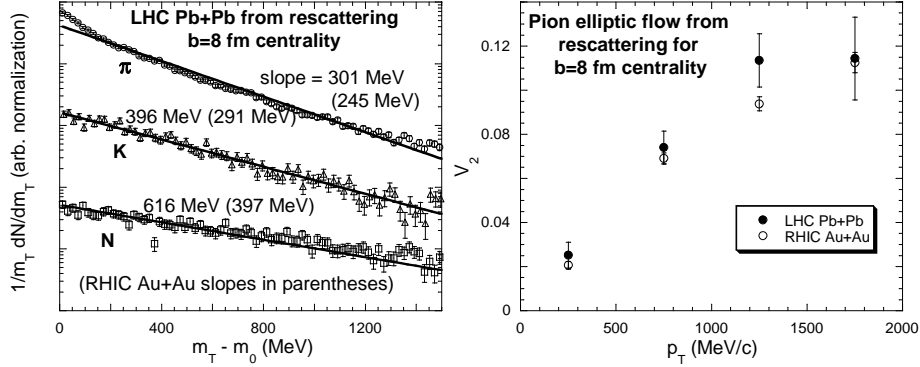}}
\caption{$m_T$ distribution and pion elliptic flow predictions for
LHC Pb+Pb compared with RHIC Au+Au for $b=8$ fm centrality
collisions at mid-rapidity from HRM. Note that in the plot on the
left, corresponding RHIC slope parameters from HRM are show in
parentheses for comparison with the LHC values.} \label{fig:hbt2}
\end{figure}

Figure 30 shows the radial and elliptic flow predictions for LHC
Pb+Pb compared with RHIC Au+Au from HRM. As seen in the
$m_T$-distribution plot on the left, although all species of
particles start from a common temperature in the calculation, after
rescattering the exponential slope parameters follow the usual
radial flow pattern of $slope(\pi)<slope(K)<slope(N)$ for both LHC
and RHIC. The slopes are seen to be consistently larger at LHC than
RHIC, as well as for the degree of radial flow which is built up.
Looking at the plot of pion elliptic flow vs. $p_T$ on the right, it
is seen that LHC and RHIC give about the same values. This is due to
the elliptic flow stabilizing at a very early stage in the HRM
calculation.

\begin{figure}
\hspace{.5in} \scalebox{1.}{\includegraphics{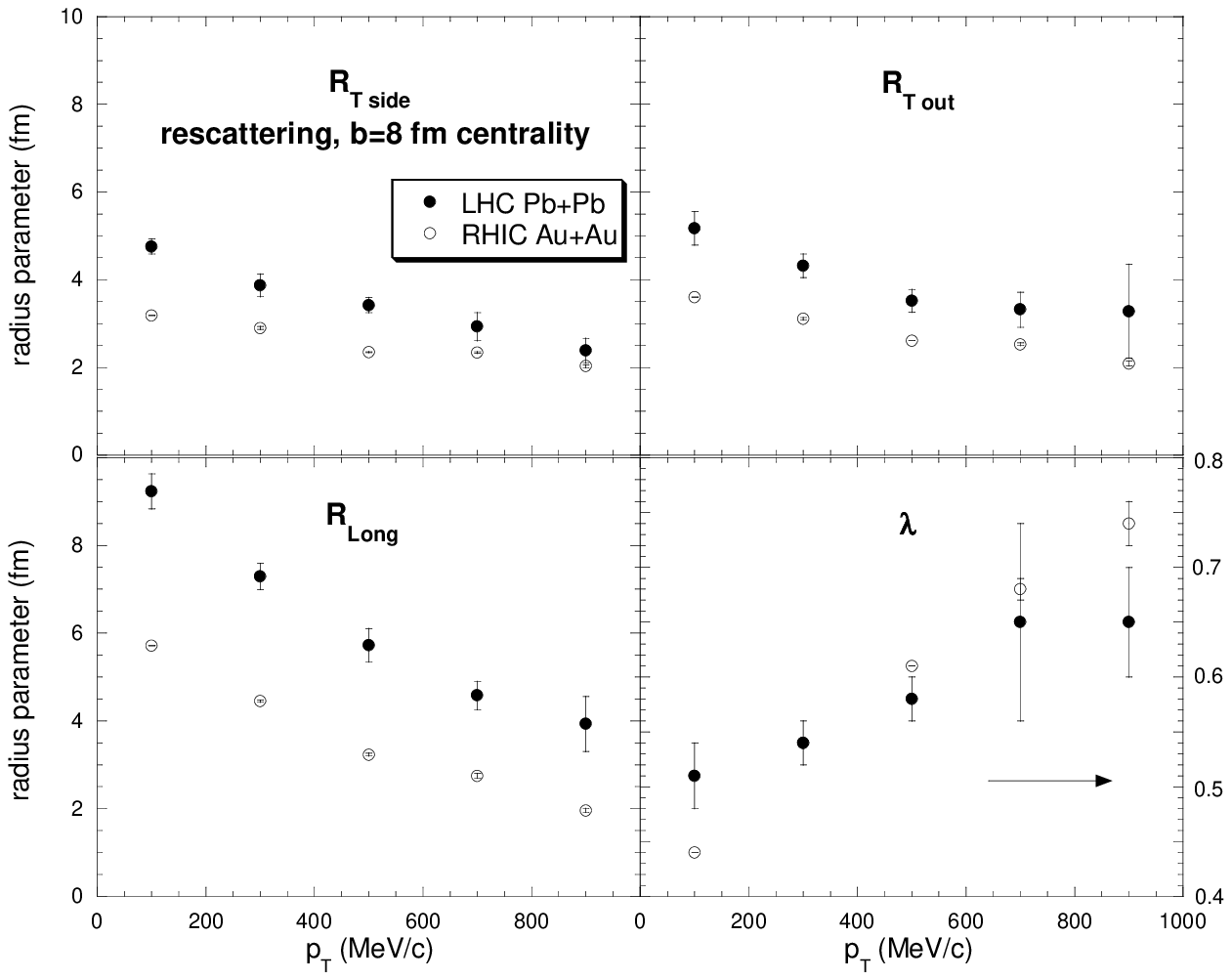}}
\caption{Pion HBT predictions for LHC Pb+Pb compared with RHIC Au+Au
for $b=8$ fm centrality collisions at mid-rapidity from HRM.}
\label{fig:bh3}
\end{figure}

Figure 31 shows the pion HBT parameters vs. $p_T$ for LHC Pb+Pb
compared with RHIC Au+Au from HRM. The transverse radius parameters,
$R_{T side}$ and $R_{T out}$, are seen to be somewhat larger and
show a stronger $p_T$ dependence for LHC as compared with RHIC. The
longitudinal radius parameter, $R_{Long}$ is seen to be
significantly larger for LHC as compared with RHIC, clearly
reflecting that the pion freezeout times at LHC are twice as long as
at RHIC according to HRM. The $\lambda$ parameter is seen to
increase with increasing $p_T$ in the same way for both LHC and
RHIC, reflected the reduced influence of long-lived resonances at
the higher $p_T$ values.

Summarizing the results of this preliminary study, it is predicted
from HRM that medium-peripheral ($b=8$ fm) LHC Pb+Pb collisions will
produce more radial flow and larger HBT radii than the analogous
RHIC Au+Au collisions, although elliptic flow and the $\lambda$
parameter values will look the same.

\section{Summary}
Over the past 25 years the HBT interferometry technique as applied
to relativistic heavy ion collisions has evolved from being little
more than a curiosity to becoming a standard tool applied by most
RHI experiments to help understand their data. It has evolved in
both its level of theoretical sophistication as well as in the level
of precision at which HBT measurements can be carried out. The issue
still remains of how to physically interpret the HBT observables
which are measured. As is the clear theme of the present review,
hadronic scattering models provide a means of addressing this issue
of interpretation. Because of their success in describing the
behavior of experimental HBT parameters, in addition to other
measured observables such as the radial and elliptic flow, such
models can be used in conjunction with experiments to view
relativistic heavy ion collision at an earlier stage prior to the
randomizing effects of the rescattering process. It will be more
than interesting to carry out HBT measurements at the LHC in several
years and thus to see if nature presents us with a host of new HBT
puzzles to solve.

\section*{Acknowledgment}
This work was supported by the U.S. National Science Foundation
under grant PHY-0355007.


\begin{thebibliography}{0}
\bibitem{cris2}
Hanbury-Brown mentioned his measurement of Alpha Centauri during a coffee break
at the CRIS'98 conference. He said that he was surprised that the HBT signal was
half of what he expected from observing its brightness with other methods (e.g.
a single photomultiplier tube). At first, he thought there was something wrong
with the detector setup, but finally convinced himself that a binary star should
give half the expected HBT signal. He assigned the derivation of this to us as a
homework problem, and this is my solution to it.
\end{thebibliography}
\end{document}